\documentclass[aps,prb,showpacs,twocolumn,floats,epsfig,pdflatex]{revtex4}
\usepackage{amssymb}
\usepackage{amsbsy}
\usepackage{amsmath}
\usepackage{epsfig}
\usepackage{epstopdf}
\usepackage{dcolumn}
\usepackage{graphicx}
\usepackage{epsfig,amssymb,amsmath}
\usepackage{color}
\begin{document}
\title{Josephson current between two $p$-wave superconducting nanowires in the presence of Rashba spin-orbit interaction
and Zeeman magnetic fields}
\author{E. Nakhmedov$^{1,2}$, B. D. Suleymanli$^3$, O. Z. Alekperov$^{1}$, F. Tatardar$^{1,4}$, H. Mammadov$^2$, A. A. Konovko$^{5}$,
A. M. Saletsky$^{5}$, Yu. M. Shukrinov$^{6,7}$, K. Sengupta$^{8}$, and B. Tanatar$^{9}$}
\affiliation{
$^{1}$Institute of Physics of National Academy of Sciences of Azerbaijan, H. Javid ave. 133, Baku, AZ-1143, Azerbaijan,\\
$^{2}$ Baku Engineering University, Hasan Aliyev str. 120, Absheron AZ-0101 , Baku, Azerbaijan, \\
$^{3}$ National Research Nuclear University MEPhI, 115409 Moscow, Russian Federation\\
$^{4}$Khazar University, Mahsati str. 41, AZ 1096, Baku, Azerbaijan,\\
$^{5}$ Faculty of Physics, Moscow State University, Leninskie Gory 1-2, 119991 Moscow, Russian Federation\\
$^{6}$ BLTP, JINR, Dubna, Moscow region, 141980, Russian Federation,\\
$^{7}$ Dubna State University, Dubna, Moscow region, 141980, Russian Federation,\\
$^{8}$ School of Physical Sciences, Indian Association for the Cultivation of Science, Jadavpur, Kolkata 700 032, India,\\
$^{9}$ Physics Department, Bilkent University, 06800 Ankara, Turkey}
\date{\today}

\begin{abstract}
Josephson current between two one-dimensional nanowires with
proximity induced $p$-wave  superconducting pairing is calculated in
the presence of Rashba spin-orbit interaction, in-plane and normal
magnetic fields. We show that Andreev retro-tunneling is realized by
means of three channels. The main contribution to the Josephson
current gives a scattering in a conventional particle-hole channel,
when an electron-like quasiparticle reflects to a hole-like
quasiparticle with opposite spin yielding a current which depends
only on the order parameters' phase differences $\varphi$ and
oscillates with $4\pi$ period. Second anomalous particle-hole
channel, corresponding to the Andreev reflection of an incident
electron-like quasiparticle to an hole-like quasiparticle with the
same spin orientation, survives only in the presence of the in-plane
magnetic field. The contribution of this channel to the Josephson
current oscillates with $4\pi$ period not only with $\varphi$ but
also with orientational angle of the in-plane magnetic field
$\theta$ resulting in a magneto-Josephson effect. Third anomalous
particle-particle channel, which represents a reflection of an
electron-like (hole-like) quasiparticle to a electron-like
(hole-like) quasiparticle with opposite spin-orientation, oscillates
only with the in-plane magnetic field orientation angle $\theta$. We
present a detailed theoretical analysis of both DC and AC Josephson
effects in such a system showing contributions from all these
channels and discuss experiments which can test our theory.

\end{abstract}
\maketitle
\section{Introduction}
\label{intro}

A key model for realization of Majorana fermion (MF) in a condensed
matter is a spinless $p$-wave superconductor (SC) \cite{kitaev01,
kitaev03}. Majorana zero modes are excitations at zero energy which
are typically localized at interface of the topological-non-topological
phases and spatially separated from one another. They emerge as
electrically neutral fermions indistinguishable from their
antiparticles in subgap quasiparticle excitation spectrum of a
topological superconductor (TSC). In $s$-wave superconductors,
the quasiparticle excitations at the top of the supercoducting gap
are indeed equal coherent superpositions of electrons and holes
with opposite spins, and thus electrically neutral. Nevertheless,
such a superposition of fermionic quasiparticles is not
self-conjugate due to existence of spin. Therefore, MF can not
appear in $s$-wave SC, and it is expected to occur in an effectively
spinless $p$-wave SC \cite{kitaev01}. In $p$-wave SC model odd
number of Majoranas reside at each end of the superconductor.
However, electrons in conventional materials have spin-half
particles; thus, the notion of a spinless SC does not seem
immediately relevant to real physical systems.

\begin{figure}[t]
\resizebox{.50\textwidth}{!}{%
\includegraphics[width=1.0cm]{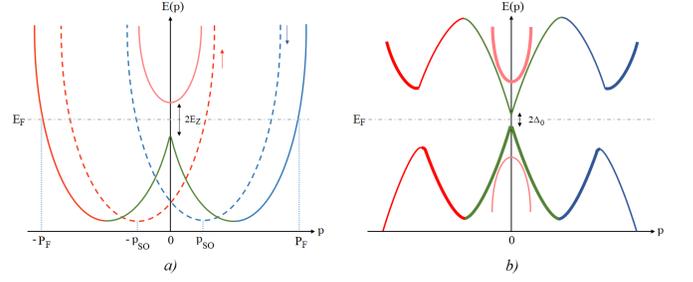}}
\caption {(a) Splitting of an electron's energy spectrum in the presence of Rashba spin-orbital interaction
is depicted by dashed curve. Zeeman magnetic field removes the degeneracy at the cross-point of two spin-
polarized spectra, and opens a gap of $2E_z$ width; (b) the band structure of 1D semiconductor with strong spin-
orbit interaction, transformed to a superconducting state due to proximity effect. }
\label{Fig1}
\end{figure}

Recently it was suggested that a topological (spinless $p$-wave)
superconductivity can be effectively realized either in a
spin-polarized normal metal or in a semiconductor nanowire with
strong spin-orbit coupling under Zeeman magnetic field
proximity-coupled to a conventional spin-singlet ($s$-wave) bulk SC
\cite{lsd10, oro10}. Spin-orbit interactions split the energy
spectrum shifting the energy branches along the momentum axis.
In contrast, an in-plane Zeeman magnetic field
shifts the energy of up- and down-spin electrons opening a gap in
the spectrum at zero momentum. When the chemical potential is
located in the gap the upper spin-subband becomes empty, and the
system is transformed to an effectively ``spinless'' electron model.
However, owing to the spin-orbit interactions-induced rotation of the spins at
the opposite Fermi points, the proximity-induced $s$-wave
interaction opens a pairing gap in the spectrum. The resulting state
is closely related to a spinless $p$-wave superconductor. One of the
virtues of this model is that the proximitized nanowire can be
driven into a topological phase by tuning the magnetic field or the
chemical potential (Fig. \ref{Fig1}). The emergence of Majorana zero
modes at a certain critical value of a control parameter is
necessarily accompanied by the closing of the bulk gap
\cite{kitaev01}, which corresponds to a topological quantum phase
transition (that is, a quantum phase transition between
topologically trivial and non-trivial states).

The topological phase in these wires is stable with
respect to small perturbations (such as disorder) as long as they do
not cause the bulk gap in the spectrum to collapse. The ability to
realize the topological phase depends on the effective spin-orbit
energy $E_{so}$, the proximity-induced gap $\Delta$ and the
effective Zeeman energy $V_Z$ in the heterostructure. Note that
highly controlled zero-energy quasi-particles Majoranas, produced in
thin wire, can be utilized as a quantum information carrier qubit in
quantum computer technology. MFs are exotic non-Abelian fermions
obeying non-Abelian braiding statistics \cite{rg00, ivanov01,
aoro11}. This unique property makes MF ideal for fault-tolerant
quantum computation. These MFs typically arise at defect sites as
Abrikosov vortex cores in bulk SCs, at interface of dielectric/TSC or
normal metal/TSC, or at edges of TSC as localized excitations, and are
topologically protected against any local perturbations.
Essentially, two key points for the emergence of MFs are presented
here: spin-orbit coupling (SOC) and superconducting proximity
effect. A neutral excitation in a superconductor has a special
property owing to the inherent particle-hole symmetry of the
material: it is bound to zero energy, so that there is no cost to
occupy such a state. One dimensional (1D)
topological superconductor supports a non-local fermionic mode
comprising two Majorana zero modes localized at opposite ends of the
chain and are separated by a distance that can be much larger than
the superconducting coherence length. An odd number of Majorana zero
modes emerges per wire end; this is consistent with
the fact that an even number of Majorana modes can pair up and
locally form an Andreev state (which is a conventional fermionic
state).

Recent investigations have shown \cite{gr01, ljl14, rm15, yw16} that
a proximity of semiconductor with $s$-wave superconductor induces
not only $s$-wave but also $p$-wave superconducting pairing in the
semiconductor. The pairing symmetry of a BCS-type two dimensional
(2D) superconductor without inversion symmetry when the twofold
degeneracy of the electron energy spectrum is lifted by SOC has
been studied \cite{gr01, edelstein89, faks04} to be a mixture
of singlet and triplet symmetries.
Recently Reeg and Maslov have shown \cite{rm15} by directly solving
the fully quantum-mechanical Gor'kov equations that spin-triplet
superconducting correlations are induced by Rashba SOC in both 1D
and 2D proximity junctions via the proximity effect. Furthermore,
the induced triplet component in 1D was shown to vanish when
integrated over the momentum; this result is in agreement with Ref.\
[\onlinecite{ljl14}]. The induced triplet amplitude  in 2D was found
to have an odd-frequency component that is isotropic in momentum.

In this work we study Josephson junction ($JJ$) of two superconductors with
$p$-wave pairing of spinfull electrons separated by a
$\delta$-function like insulator potential. $JJ$
consisting of $p$-wave superconductors in both sides of the junction
has been studied \cite{ksy04} for a $\delta$-like insulator
potential in the absence of SOC and Zeeman magnetic field. Since the
superconductor is in the topological phase, a fractional oscillation
of Josephson current was obtained in Ref.\ [\onlinecite{ksy04}]. The
emergence of Majorana zero mode results in exotic Josephson effects
\cite{kitaev01, fk09, jpar11, jpar13, nats17}, when the current
flowing between two topological superconductors in the junction
oscillates with a fractional periodicity $4 \pi$ instead of $2\pi$
periodicity in a conventional Josephson junction. Additionally,
“spin Josephson current” \cite{nb04, asano06, brydon09, bat11}
may flow across the junction, which is shown \cite{jpar11, jpar13}
also to be $4\pi$ periodic in the field orientations as a
manifestation of the Majorana modes.

Josephson current in a junction of two $s$-wave symmetric superconductors
has been studied by us in our previous work \cite{nats17}. In this work we study
the Josephson current through a junction of two superconductors with $p$-wave pairing of
spinfull electrons. This problem has been usually studied for spinless model \cite{kitaev01},
although many aspects of spinfull $p$-wave symmetric $JJ$ have been investigated
\cite{zclw12, kecs14, ehm16, bs19,mkc19} in the literatures. Furthermore, in this paper we
want to understand how do SOC and external magnetic fields change
Josephson current in p-type $JJ$ separated by a
$\delta$-potential insulator. Similar effects have studied in Ref.\
[\onlinecite{nats17}] for a junction consisting of $s$-wave SCs
separated by $\delta$-potential thin insulator. Note that, the Josephson
current in the case of proximity induced both $s$- and $p$-wave pairings
in $JJ$, as argued in several recent papers \cite{gr01, ljl14, rm15, yw16},
can be found by summing up a corresponding result of Ref. [\onlinecite{nats17}]
with that presented in this paper. We show in this paper
that the Andreev tunneling occurs in three channels, and clarify the
origin of these channels. Two additional channels seem to vanish with
{\it in-plane} magnetic field. All contributions to the Josphson current
oscillate with fractional periodicity either with phase difference of the
SC order parameters $\varphi$ or with tilded angle $\theta$ between the
in-plane magnetic field and JJ. Simultaneous action of SOC and Zeeman
magnetic fields results in openning of a forbidden gap in dependence
of Josephson current on the phases at some definite values of SOC and
magnetic fields.

The paper is organized as
follow. In the next Section formulation of the problem is presented.
In Section III Andreev bound state energy is calculated for
different values of the external parameters as SOC constant, Zeenam
energies for the magnetic fields normal to the junction $h$ and the
tilted magnetic field ${\bf B}$ lying in the junction plane and
forming an angle $\theta$ with SC wire. Section IV describes
Josephson current as well as magneto-Josephson effect in the
junction. $ac$-Josephson current and effects of SOC and magnetic field
on the Shapiro step in the $JJ$ of p-wave superconductors  are studied
in the last Section V.

\section{Model and formulation of the problem}
\label{sec2}
We consider a junction of two $1D$ nanowires of proximity induced
$p_x$-wave pairing symmetry superconductivity, having the effective pairing potentials
$\Delta_L$ and $\Delta_R$ on the left(L)- and right (R)-side of an insulating potential barrier
separated two superconductors, in the presence of Rashba spin-orbital interaction and external
Zeeman magnetic fields. Hamiltonian for such a system reads
\begin{equation}
\hat{H}= \hat{H}_{SC} + \hat{H}_R, \label{H}
\end{equation}
where $\hat{H}_{SC}$ is Hamiltonian of the nanowire in the presence
of external magnetic fields and $\hat H_R$ represents Rashba SOI.
The former term is given by
\begin{eqnarray}
\hat{H}_{SC} &=& \int dx \Bigg\{  \sum_{\sigma, \sigma'}
\psi_{\sigma}^{\dag}(x) \Big[[\xi_{\hat k}+U(x)] \sigma_0 + h
\sigma_z \nonumber\\
&& + B \{[\sigma_x \cos \theta_1 + \sigma_y \sin \theta_1] \theta(-x) +
[\sigma_x \cos \theta_2 \nonumber\\
&& + \sigma_y \sin \theta_2]
\theta(x) \} \Big] \psi_{\sigma'}(x) \nonumber\\
&& +  \left[\Delta_L \theta(-x) + \Delta_R \theta(x)\right]
\psi_{\uparrow}^{\dag}(x) \psi_{\downarrow}^{\dag}(x)+ {\rm h.c.}
\Bigg\}, \label{H-sc}
\end{eqnarray}
where $\xi_{\hat k} = \epsilon
\left(\frac{\hbar}{i}\frac{\partial}{\partial x}\right) -\epsilon_F$
denotes the electron kinetic energy as measured from the Fermi
level $\epsilon_F$, $\psi_{\sigma}(x)$ is the electron annihilation
operator, $h$ and ${\bf B}$ are external Zeeman magnetic fields in
$z$ direction and in the {\it x-y} plane respectively, $\theta(x)$ is
the Heaviside step function, and $\sigma_{x,y,z}$ and $\sigma_0$
denote Pauli and identity matrices respectively in spin space. Note
that the magnetic field ${\bf B}$ forms an angle $\theta$ with wire
which can be tuned externally. In what follows, we choose ${\bf B}$
in the left side of the junction to be aligned along the wire
($\theta_L=0$) while in the right side it is chosen to make an angle
$\theta$ with it ($\theta_R= \theta$). In Eq.\ (\ref{H-sc}), the pairing
potential $\Delta_R$ in the right of the junction is chosen to have
a phase difference $\varphi$ compared to its left counterpart:
$\Delta_R = |\Delta| \exp(i\varphi)$ and $\Delta_L = |\Delta|$.  The
potential $U(x)=U_0 \delta (x)$, located at $x=0$, represents
the barrier potential between two superconductors. The
Hamiltonian of Rashba SOI can be written as
\begin{eqnarray}
\hat{H}_{R}= \sum_{\sigma, \sigma'} \int dx \psi_{\sigma}^{\dag}(x) \alpha \left[v_x \sigma_z\right]\psi_{\sigma'}(x),
\label{rashba-o}
\end{eqnarray}
where $\alpha$ is the strength of Rashba SOI which is chosen to be
the same for both wires.

The order parameter $\Delta_{\alpha, \beta}^t (\hat{\bf k})$ for triplet pairing with $S=1$ can be presented
as \cite{leggett75, ms98}
\begin{equation}
\Delta_{\alpha, \beta}^t (\hat{\bf k})= \Delta {\bf d}({\bf k})(i {\bf \sigma} \sigma_y)_{\alpha, \beta}
\end{equation}
where ${\bf d}({\bf k})$ is odd function of ${\bf k}$, ${\bf d}({\bf k})=-{\bf d}(-{\bf k})$, and can be expanded
over spherical harmonics
\begin{equation}
d_{\alpha}({\bf k})=\sum_{m=-l}^{l}b_{l,m}^{\alpha}Y_{l,m}(\hat{\bf k}).
\label{d}
\end{equation}
The quantum number $l$ in Eq. (\ref{d}) takes odd values $1, 3, \dots$ corresponding to states of $p, f, \dots$-
pairings. The coefficient $b_{l,m}^{\alpha}$ can be identified as $\alpha$ component of the superconducting order
parameter with given $l$ and $m$. For a simple case of $p$-wave pairing for $l=1$ and $m=-1, 0,1,$
$d_{\alpha}({\bf k})$ can be expressed with the appropriate expressions for the spherical harmonics
$Y_{1,-1} \propto (\hat{k}_x-i\hat{k}_y)$, $Y_{1,0} \propto \hat{k}_z$, and $Y_{1,1} \propto (\hat{k}_x+i\hat{k}_y)$, as,
\begin{equation}
d_{\alpha}({\bf k})=A_{\alpha, i}\hat{k}_i.
\end{equation}
Thereby, the $p_x$-wave symmetric order parameter $\Delta_{a, \sigma}(x, k_x)$  on the $a=R, L$-side of
a junction between two superconductors aligned along the $x$-axis can be expressed as,
\begin{equation}
\Delta_{a, \sigma}(x, k_x)=\Delta_a \frac{k_x}{k_F}.
\label{px}
\end{equation}

It is advantageous to use a four component field operator for bulk
superconductor at the right ($a=R$) and left ($a=L$) side of the junction as,
\begin{eqnarray}
\Psi^{\dag}_a(x)=\left(\psi_{a,\uparrow, +}^{\dag}(x), \psi_{a,
\downarrow,+}^{\dag}(x), \psi_{a, \downarrow, -} (x), \psi_{a,
\uparrow, -}(x)\right) \label{op1}
\end{eqnarray}
Here the third subscript of the annihilation operator (which we
shall designate henceforth as $b$) labels the right- ($b=+$) and the
left-moving $(b=-$) quasiparticles respectively while the index
$a=R,L$ denotes either right ($R=-$) or left ($L=+$) superconductor.
In terms of the field operator given by Eq.\ (\ref{op1}), the
Hamiltonian (Eq.\ (\ref{H})) can be written as $\hat{H}= \sum_{a=R,L}
\int dx \Psi^{\dag}_a(x) \mathcal{H}_a \Psi_a(x)$ using the Pauli
matrices $\sigma_i$ in spin- and $\tau_i$ in particle-hole spaces.
From Eqs. (\ref{H}) and (\ref{H-sc}), we find
\begin{eqnarray}
&& \mathcal{H}_R=\xi_{k, b} \tau_z \sigma_0 + h \tau_0 \sigma_z -i k
\alpha  \tau_z \sigma_z \label{H2} \\
&& + B \tau_z \left(\sigma_x \cos \theta + \sigma_y \sin \theta  \right)
+|\Delta| (\tau_x \cos \varphi - \tau_y \sin \varphi) \sigma_z,
\nonumber
\end{eqnarray}
and ${\mathcal H}_L= {\mathcal H}_R (\theta=0;\varphi=0)$. In Eq.\
(\ref{H2}), the energy spectrum of the electrons are linearized
around the positive and negative Fermi momenta leading to $\xi_{k,
b}= b v_F\left(-i \frac{\partial}{\partial x} - k_F\right)$, where
$v_F$ is the Fermi energy. Note that the Hamiltonian $\mathcal
{H}_{R,L}$ acquires a magnetism-superconductivity duality
\cite{nab08, jpar13} in the absence of the kinetic term, implying
that it becomes invariant under the transformation $\{\Delta,
\epsilon_F, \varphi, \tau_i \} \to \{B, h, - \theta, \sigma_i \}$. The
existence of a magneto-Josephson effect in a topological insulator
is known to be a result of this duality \cite{jpar13}.
We shall see that for the system we study, the
magneto-Josephson effect takes place even in the presence of the
additional quadratic kinetic energy term of the electrons.

The energy spectrum of a quasi-particle in  a 'bulk' quasi-1D superconductor is
determined from the expression ${\rm Det} |\mathcal{H}_R - E|=0$, yielding
\begin{widetext}
\begin{equation}
\left(E^2 -v_F^2k^2 + \alpha^2 k^2 -h^2-B^2 -|\Delta|^2 \right)^2 +4\left(Ekv_F + h \alpha k \right)^2 -
4 |\Delta|^2 \left(v_F^2 k^2 + B^2 + h^2 \right)=0.
\label{Eo}
\end{equation}
\end{widetext}
This equation does not yield a simple analytic expression for $E$, while it contains a
linear in energy term, which is a result of an alignment of ${\bf h}$ and the effective
magnetic field of the SOI $\propto \alpha k$. The linear in $E$ term vanishes for either
$h=0$ or $\alpha =0$, and Eq. (\ref{Eo}) turns to quadratic equation for $E^2$, which
gives two symmetric dispersion branches for quasi-particles and quasi-holes. Note that square
of the momentum $k^2_{\pm}$, where the subscripts indices indicate the spin branches, can be
obtained from Eq. (\ref{Eo}). The evident expression for $k^2_{\pm}$ is presented by Eq. (\ref{k2}) in
Appendix.
Equation (\ref{Eo}) is strongly simplified for different limiting cases, and yields the
following expressions for the energy dispersion,
\nopagebreak
\begin{widetext}
\begin{equation}
E=
\begin{cases}
\pm \sqrt{|\Delta|^2 -v_F^2 k^2}  & \text{if} \qquad  \alpha = B =h=0,
\\
s\left[\sqrt{|\Delta|^2 - v_F^2 k^2} \pm  \sqrt{ B^2 + h^2}\right]    & \text{if} \qquad \alpha =0, B \neq 0, h \neq 0,
\\
s\sqrt{B^2+|\Delta|^2-(v_F^2+\alpha^2)k^2 \pm 2\sqrt{v_F^2\alpha^2k^4+|\Delta|^2B^2-v_F^2k^2B^2}} & \text{if} \qquad
h=0, \alpha \neq 0, B \neq 0,
\label{E4}
\end{cases}
\end{equation}
\end{widetext}
where $s=\pm$ indicates the particle- and hole-branches of the spectrum.
The energy levels of the Bogolyubov-de Gennes  (BdG) quasi-particles lie in the gap,
symmetric to the Fermi level. SOI and/or magnetic field $h$ split both electron and
hole levels due to Rashba 'momentum-shifting' and/or Zeeman effect. The 'Fermi
points' around $+k_F$ and $-k_F$ are split also due to these
effects. At the same time, the magnetic field makes the energy dispersion asymmetric.
Note that in our case, all energies are measured from the Fermi energy;
thus the condition for realization of a topological non-trivial superconducting
gapped phase is $|\Delta|^2 \ge B^2+h^2$, \cite{jpar13}. Indeed, zero energy mode ($E=0$) at
the center of the Brillouin zone ($k=0$) appears according to Eq. (\ref{Eo}) under the
condition $|\Delta|^2-B^2 - h^2=0$.

BdG equations for an isolated 'bulk' superconductor
in the case of an infinitely high potential between the right (R)- and left (L)
parts ($a=R, L$) of the barrier is written as
\begin{equation}
\mathcal{H}_a \mathbf{\eta}_a(x)= E \mathbf{\eta}_a(x), \quad a= R,
L\label{Sch}
\end{equation}
where the four-component vector $\eta_a(x)=\left(\eta_{a,\uparrow, +}^{\dag}(x), \eta_{a,
\downarrow,+}^{\dag}(x), \eta_{a, \downarrow, -} (x), \eta_{a,
\uparrow, -}(x)\right)$ denotes the BdG wave function.
In order to get the explicit expressions for the wave functions
$\eta_{a,\sigma,b}$ and $\eta_{a,{\bar \sigma},b}^{\ast}$  we write
Eq.\ (\ref{Sch}) for finite value of the external parameters
${\bf B}$, $h$ and $\alpha$ as
\begin{eqnarray}
&& (E + i a b v_F k + iab \alpha k - h)\eta_{a, \uparrow, b} -
Be^{-i\theta} \eta_{a, \downarrow, b}  \nonumber\\
&&- \Delta_a \eta^{\ast}_{a, \downarrow, \bar{b}}= 0 \label{Sch1}\\
&&(E + i a b v_F k - i ab \alpha k + h)\eta_{a, \downarrow, b} -
Be^{i \theta} \eta_{a, \uparrow, b} \nonumber\\
&& + \Delta_a \eta^{\ast}_{a, \uparrow, \bar{b}} = 0 \label{Sch2}\\
&& (E - i a b v_F k - i ab \alpha k - h)\eta^{\ast}_{a, \downarrow,
\bar{b}}+ Be^{-i \theta} \eta^{\ast}_{a, \uparrow, \bar{b}}  \nonumber\\
&&- \Delta_a^{\ast} \eta_{a, \uparrow, b} = 0 \label{Sch3}\\
&& (E - i a b v_F k + i a b \alpha k + h)\eta^{\ast}_{a, \uparrow,
\bar{b}} + Be^{i \theta}
\eta^{\ast}_{a, \downarrow, \bar{b}} \nonumber\\
&&+\Delta_a^{\ast} \eta_{a, \downarrow, b}= 0. \label{Sch4}
\end{eqnarray}
In order to understand the features of Eqs. (\ref{Sch1})-(\ref{Sch4})
one considers several asymptotic cases. In the absence of the in-plane
magnetic field ${\bf B}=0$, these equations link a particle wave
function $\eta_{a, \sigma, b}$ with the hole one $\eta^{\ast}_{a, \bar{\sigma}, \bar{b}}$
of an opposite spin-polarized and opposite direction-moved quasiparticle
state \cite{ksy04} and vice-versa, provided that the system is
in a superconducting phase, $\Delta_a \neq 0$. Instead, in the absence of
a superconducting phase, $\Delta_a=0$, these equations link a particle (hole)
wave function $\eta_{a, \sigma, b}$ with the particle (hole) wave function with opposite
spin-polarized quasiparticle state $\eta_{a, \bar{\sigma}, b}$ moving in the
same direction provided that ${\bf B} \neq 0$. However, in the presence of in-plane magnetic field ${\bf B} \neq 0$
in the superconduction phase $\Delta_a \neq 0$, Eqs. (\ref{Sch1})-(\ref{Sch4}) connect
a particle wave function $\eta_{a, \sigma, b}$ (a hole wave function $\eta^{\ast}_{a, \sigma, \bar{b}}$)
with hole (particle) wave functions with the same $\eta^{\ast}_{a, \sigma, \bar{b}}$ ( $\eta_{a, \sigma, b}$)
and opposite  $\eta^{\ast}_{a, \bar{\sigma}, \bar{b}}$ ( $\eta_{a, \bar{\sigma}, b}$)  spin-polarized
quasiparticle states moving in the direction opposite to the particle (hole) one.
Eqs. (\ref{Sch1})-(\ref{Sch4}) allow us to calculate all
possible ratios $\eta^{\ast}_{a, \sigma, b}/\eta_{a,{\bar
\sigma},{\bar b}}$, $\eta^{\ast}_{a, \sigma, b}/\eta_{a,\sigma,
{\bar b}}$, and $\eta_{a, \sigma,b}/\eta_{a,{\bar \sigma}, b}$,
$\eta^{\ast}_{a, \sigma,b}/\eta^{\ast}_{a,{\bar \sigma}, b}$.
Furthermore, we note that only the ratio $\eta^{\ast}_{a, \sigma,
b}/\eta_{a,{\bar \sigma},{\bar b}}$ is non-zero for ${\bf B}=0$, which corresponds
to the conventional Andreev reflection at the boundary of a superconductor
with normal metal or insulator \cite{ksy04}.
Eqs. (\ref{Sch1})-(\ref{Sch4}) provide the following expressions for
these ratios for arbitrary values of the parameters $\alpha$, ${\bf B}$, $h$,
\begin{eqnarray}
\hspace{-10mm}&&\frac{\eta^{\ast}_{a, \uparrow, {\bar b}}}{\eta_{a, \downarrow, b}}= -\frac{1}{\Delta_{a, b}}\Bigg\{E+iabkv_F-iab\alpha k+h-
\nonumber\\
\hspace{-10mm}&&\frac{2B^2(E - iab\alpha k)}{(E-h)^2+(kv_F+\alpha k)^2 + B^2-\Delta_{a, b}^2}\Bigg\}
\label{ratup-down}\\
\hspace{-10mm}&&\frac{\eta^{\ast}_{a, \uparrow, {\bar b}}}{\eta_{a, \uparrow, b}}=\frac{B e^{i \theta}}{\Delta_{a, b}}\Bigg\{1-\nonumber\\
\hspace{-10mm}&&\frac{2(E+iab\alpha k)(E+iabkv_F-iab\alpha k + h)}{(E+h)^2+(kv_F-\alpha k)^2 + B^2-\Delta_{a, b}^2}\Bigg\},
\label{ratup-up}\\
\hspace{-10mm}&&\frac{\eta_{a, \uparrow, b}}{\eta_{a, \downarrow, b}}= \frac{2B (E-i ab\alpha k)e^{-i \theta}}{(E-h)^2+(kv_F+k \alpha)^2 +
B^2 -\Delta_{a, b}^2};
\label{ratup-down2}
\end{eqnarray}
where $\Delta_{a, b}=b \Delta_a$ according to Eq. (\ref{px}), and $\Delta_R=|\Delta| \exp(i \varphi)$
and $\Delta_L=|\Delta|$. Note that the expressions for $\frac{\eta^{\ast}_{a, \downarrow, {\bar b}}}{\eta_{a, \uparrow, b}}$,
$\frac{\eta^{\ast}_{a, \downarrow, {\bar b}}}{\eta_{a, \downarrow, b}}$, and $\frac{\eta_{a, \downarrow, b}}{\eta_{a, \uparrow, b}}$
can be obtained respectively from Eqs. (\ref{ratup-down}), (\ref{ratup-up}), and (\ref{ratup-down2})
by replacing $\theta \to - \theta$, $\alpha \to  - \alpha$, $h \to -h$, and by reversing the total
sign of these expressions. According to these expressions the reflection channels, determined by the ratios
$\frac{\eta^{\ast}_{a, \downarrow, {\bar b}}}{\eta_{a, \downarrow, b}}$, and $\frac{\eta_{a, \downarrow, b}}{\eta_{a, \uparrow, b}}$,
vanish with in-plane magnetic field ${\bf B}$.

We note from Eqs. (\ref{Sch1})-(\ref{Sch4}) that the
dependencies of these equations on $\varphi$ and $\theta$ are
completely removed by transforming the wave functions as
\begin{eqnarray}
\eta^{\ast}_a(x) &\to& \left(e^{-i(\varphi-
\theta)/2}\eta_{a,\uparrow, \bar{b}}^{\ast}(x), ~e^{-i(\varphi
+\theta/2)} \eta_{a,
\downarrow,\bar{b}}^{\ast}(x), \right. \nonumber\\
&& \left.~ e^{i(\varphi+ \theta)/2}\eta_{a, \downarrow, b} (x),
~e^{i(\varphi- \theta)/2} \eta_{a, \uparrow, b}(x)\right).
\end{eqnarray}
In the transformed basis one has
\begin{eqnarray}
\frac{\eta^{\ast}_{a, \uparrow, b}}{\eta_{a,\uparrow,{\bar b}}} \to
e^{-i(\varphi - \theta)} \frac{\eta^{\ast}_{a, \uparrow,
b}}{\eta_{a,\uparrow, {\bar b}}}, \quad \frac{\eta^{\ast}_{a,
\downarrow, b}}{\eta_{a,\downarrow, {\bar b}}} \to e^{-i(\varphi +
\theta)} \frac{\eta^{\ast}_{a, \downarrow,
b}}{\eta_{a,\downarrow, {\bar b}}} \label{dagup-up} \\
\frac{\eta^{\ast}_{a, \uparrow, b}}{\eta_{a,\downarrow, {\bar b}}}
\to e^{-i\varphi} \frac{\eta^{\ast}_{a, \uparrow,
b}}{\eta_{a,\downarrow, {\bar b}}}, \quad   \frac{\eta^{\ast}_{a,
\downarrow, b}}{\eta_{a,\uparrow, {\bar b}}} \to e^{-i\varphi}
\frac{\eta^{\ast}_{a, \downarrow,
b}}{\eta_{a,\uparrow, {\bar b}}} \label{dagup-down} \\
\frac{\eta^{\ast}_{a, \uparrow, b}}{\eta_{a,\downarrow, b}^{\ast}}
\to e^{i\theta} \frac{\eta^{\ast}_{a, \uparrow,
b}}{\eta_{a,\downarrow, b}^{\ast}}, \quad  \frac{\eta_{a, \uparrow,
b}}{\eta_{a,\downarrow, b}} \to e^{-i\theta} \frac{\eta_{a, \uparrow,
b}}{\eta_{a,\downarrow, b}}. \label{up-up}
\end{eqnarray}
The different ratios that appear in the left hand-side of Eqs.\
(\ref{dagup-up})-(\ref{up-up}) can be understood on follows. The ratio
$\eta^{\ast}_{a, \sigma, b}/\eta_{a,{\bar \sigma},{\bar b}}$
corresponds to the amplitude of conventional Andreev reflection
channel which constitutes reflection of an electron-like
quasiparticle to a hole-like quasiparticle with opposite spin on a
N-S interface. In contrast, the ratio $\eta^{\ast}_{a, \sigma,
b}/\eta_{a,\sigma, {\bar b}}$ which is finite only in the presence
of SOC and/or magnetic field, represents amplitude of Andreev
reflection channel where the electron-like quasiparticle incident on
the interface is reflected to a hole-like quasiparticle state with
the same spin orientation. Finally, the ratio $\eta_{a,
\sigma,b}/\eta_{a,{\bar \sigma},b}$ represents a usual reflection
channel of an electron-like quasiparticle on the boundary without
creation of a Cooper pair in a superconducting part of the junction.
We note that the ratio of wave functions in Eq.\ (\ref{dagup-up}) depend
on both $\varphi$ and $\theta$ while those in Eqs. (\ref{dagup-down})
and (\ref{up-up}) depend on either $\varphi$ or $\theta$. This
suggests that the ratios (\ref{dagup-up}) and (\ref{dagup-down}) are
responsible for the dependence of observable parameters on the order
parameter phase difference $\varphi$, whereas the ratios
(\ref{dagup-up}) and (\ref{up-up}) are responsible for the
dependence on the  magnetic field orientation angle $\theta$.

In what follows, we shall look for the
localized subgap Andreev bound states with $\epsilon(k)< |\Delta|$
for the Josephson junction of two nanowires described by Eq.\
(\ref{H}).
\\

\section{\textbf{Andreev bound states, Josephson and magneto-Josephson effects}}
\label{sec3}
In order to obtain a solution for the Andreev
bound states for the junction described by Eq. (\ref{H-sc})
one follows the method used in Ref. [\onlinecite{ksy04}]. The energy spectrum of
an electron is splitted in the presence of Rashba SOC and/or Zeeman magnetic field,
so that the Fermi level crosses the dispersion curve at four points, corresponding to
right-mover $k_{F+}$, $k_{F-}$ and left-mover $-k_{F+}$, $-k_{F-}$ particles with
oppositely polarized spin-states (see, Fig. \ref{Fig1}), and $k_{F+}-k_{F-} \to 0$,
$k_{F+}, k_{F-} \to k_F$ as $\alpha, B, h \to 0$. Furthermore, a condensation of
the electron pairs in a superconducting state opens a gap around the Fermi level as is shown in
Fig.\ref{Fig1}b. We neglect here a difference between $k_F$ and $k_{F+}$, $k_{F-}$, and take
$k_{F+} \approx k_{F-} \equiv k_F$. We assume that a transition occurs between the states with
the same chirality. In order to obtain the wave function for $L$ and $R$ superconductors
we superpose the wave functions for the left ($-$) and right ($+$)
moving BdG quasiparticles correspondingly around the Fermi levels $k_{F}$,
and  $-k_{F}$ with arbitrary coefficients,
\begin{widetext}
\begin{eqnarray}
\mathbf{\eta}_{a}(x)= e^{{\rm sgn}(a) k x} \left\{e^{i k_F x} A_a
\left(\begin{array}{ccc}
\eta_{a,\uparrow, +}(k)  \\
\eta_{a, \downarrow, -}(k)\\
\eta_{a,\downarrow, -}^{\ast}(k)  \\
\eta_{a, \uparrow, +}^{\ast}(k)
\end{array} \right)
+e^{-i k_F  x} B_a \left(\begin{array}{ccc}
\eta_{a,\uparrow, -}(k) \\
\eta_{a, \downarrow,+}(k) \\
\eta_{a,\downarrow, +}^{\ast}(k)\\
\eta_{a, \uparrow, -}^{\ast}(k)
\end{array} \right)\right\}
\label{wave}
\end{eqnarray}
\end{widetext}
where ${\rm sgn} (a)= + (-)$ for $a=L (R)$. SOC and magnetic field remove the spin degeneracy
in a quasi-particle ($\eta_{a,b}$) and a quasi-hole ($\eta^{\ast}_{a, b}$) wave functions, and
thereby split  the wave functions written for the conventional superconductors \cite{ksy04}
as is shown in the above given expression.

Andreev bound state energies are obtained by imposing the usual boundary conditions on
each component of these wave functions  $\eta_a(x)$. For a barrier modeled by the delta function
potential $U(x)= U_0 \delta (x)$, the boundary conditions are provided at the merging
point $x=0$ of two superconductors as,
\begin{eqnarray}
\mathbf{\eta}_L(0)= \mathbf{\eta}_R(0),\qquad \qquad
\partial_x \mathbf{\eta}_R -
\partial_x \mathbf{\eta}_L= k_F Z \mathbf{\eta} (0),
\label{bc}
\end{eqnarray}
where $Z=2mU_0/\hbar^2k_F$ and the transmission coefficient $D$ is
expressed through $Z$ as $D=4/(Z^2 +4)$.

By choosing a pair of the wave functions from Eq. (\ref{wave}) and substituting they into the boundary
conditions (\ref{bc}) one gets four linear homogeneous equations. The energy of the Andreev bound state
for a transmission of the barrier through a particular channel is obtained from the determinant of
these linear homogeneous equations. Selection, e.g. the second and fourth equations of the wave function
(\ref{wave}) under the boundary conditions (\ref{bc}) yields the following expression for the determinant,
\begin{equation}
{\tilde F}_{\uparrow \downarrow}^{\ast}=
\frac{1}{D}\eta_{+,\downarrow,+}
\eta_{-,\downarrow,-}\eta_{+,\downarrow,-}\eta_{-,\downarrow,+}
F^{\ast}_{\uparrow  \downarrow}, \label{Fup-down0}
\end{equation}
where
\begin{eqnarray}
F_{\uparrow \downarrow}^{\ast}=\left[\frac{\eta^{\ast}_{-,
\uparrow,-}}{\eta_{-,\downarrow,+}} - \frac{\eta^{\ast}_{+,
\uparrow,-}}{\eta_{+,\downarrow,+}}\right]
\left[\frac{\eta^{\ast}_{+, \uparrow,+}}{\eta_{+,\downarrow,-}}
-\frac{\eta^{\ast}_{-, \uparrow,+}}{\eta_{-,\downarrow,-}}\right]-\nonumber\\
(1-D) \left[\frac{\eta^{\ast}_{+,
\uparrow,-}}{\eta_{+,\downarrow,+}} -\frac{\eta^{\ast}_{-,
\uparrow,+}}{\eta_{-,\downarrow,-}}\right]
\left[\frac{\eta^{\ast}_{-, \uparrow,-}}{\eta_{-,\downarrow,+}}
-\frac{\eta^{\ast}_{+, \uparrow,+}}{\eta_{+,\downarrow,-}}\right].
\label{Fup-down1}
\end{eqnarray}
Using Eq. (\ref{ratup-down}) in this expression one gets an explicit expression for $F_{\uparrow \downarrow}^{\ast}$
\begin{eqnarray}
&&\hspace{-5mm}F_{\uparrow \downarrow}^{\ast}=-\frac{4e^{-i\varphi}}{|\Delta|^2M^2_-}\Bigg\{ \left[(E+h) M_-+2B^2 E \right]^2 -
D \cos^2\frac{\varphi}{2} \times \nonumber\\
&& \hspace{-5mm}\left[ \left[(E+h)M_--2B^2E\right]^2 + \left[k(v_F-\alpha) M_-
+2 B^2 \alpha k\right]^2 \right] \Bigg\},
\label{Fup-down}
\end{eqnarray}
where
\begin{equation}
M_{\pm}=(E\pm h)^2 + (v_F \mp \alpha)^2 k^2  +B^2 -|\Delta|^2.
\label{M}
\end{equation}
The expression for $F_{\downarrow \uparrow}^{\ast}$ is obtained from Eq. (\ref{Fup-down}) by replacing
$\alpha \to - \alpha$, $h \to - h$ and $M_- \to M_+$, where $M_+(\alpha, h, B, \Delta)=M_-(-\alpha, -h, B, \Delta)$.
Solution of the equation $F_{\uparrow \downarrow}^{\ast}(E_{\uparrow \downarrow})=0$ for energy, where $F_{\uparrow \downarrow}^{\ast}$
is given by  Eq. (\ref{Fup-down}), yields a contribution to the Andreev overlap energy $E_{\uparrow \downarrow}$ in the
particle-hole channel.

Now we choose other pair, the first and fourth wave functions of (\ref{wave}), and substitute they into
the boundary conditions (\ref{bc}). The determinant of four linear homogeneous equations yields the
following expression to find the Andreev quasi-particle energy in the anomalous particle-hole channel, where the transition
occurs between the spin states with the same chirality,
\begin{equation}
{\tilde F}_{\sigma \sigma}^{\ast}(k) = \frac{1}{D}\eta_{+,\sigma,+}(k) \eta_{-,\sigma,-}(k) \eta_{+,\sigma,-}(k)
\eta_{-,\sigma,+}(k) F^{\ast}_{\sigma \sigma}(k),
\label{feqss}
\end{equation}
where $F^{\ast}_{\uparrow \uparrow}(k)$ ($F^{\ast}_{\downarrow \downarrow}(k)$) is obtained from Eq. (\ref{Fup-down1}) by
replacing all spin-down (all spin-up) with spin-up (spin-down). The evident expression for
$F^{\ast}_{\uparrow \uparrow}(k)$ is obtained by using the ratio (\ref{ratup-up}), which reads as,
\begin{eqnarray}
&&F_{\uparrow \uparrow}^{\ast}=\frac{16 B^2 e^{-i(\phi- \theta)}}{|\Delta|^2M^2_+}\Bigg\{ \left(E k v_F+\alpha h k\right)^2 -\nonumber\\
&&|\Delta|^2 \left(E^2+ \alpha^2 k^2 \right)\left[1-D\cos^2\frac{(\phi- \theta)}{2}\right]\Bigg\}.
\label{Fup-up}
\end{eqnarray}
The expression for $F^{\ast}_{\downarrow \downarrow}(k)$ can be obtained from Eq. (\ref{Fup-up}) by replacing
$\alpha \to - \alpha$, $h \to - h$, $\theta \to - \theta$, and $M_+ \to M_-$.
The general feature of the Andreev quasi-particle energy in the anomalous particle-hole channel $E_{\uparrow \uparrow}$
with the same spin orientation is that it takes non-zero values only in the presence of in-plane magnetic
field ${\bf B}$. Therefore, it
depends on the angle $\theta$ between the junction and in-plane magnetic field. Oscillation of the Josephson
current with $\theta$ yields a fractional magneto-Josephson effect.

Choice of the first and second equations of the wave function (\ref{wave}) under the boundary
conditions (\ref{bc}) yields the following expression to determine the Andreev bound state
energy in the anomalous particle-particle or hole-hole channel,
\begin{equation}
{\tilde F}_{\uparrow \downarrow}(k) =
\frac{1}{D}\eta_{-,\downarrow,-}(k) \eta_{+,\downarrow,+}(k)  \eta_{+,\downarrow,-}(k) \eta_{-,\downarrow,+}(k)
F_{\uparrow  \downarrow}(k),
\label{FA-up-down}
\end{equation}
where $F_{\uparrow \downarrow}(k)$ is written as
\begin{eqnarray}
F_{\uparrow \downarrow}=\left[\frac{\eta_{+,
\uparrow,+}}{\eta_{+,\downarrow,+}} - \frac{\eta_{-,
\uparrow,+}}{\eta_{-,\downarrow,+}}\right]
\left[\frac{\eta_{-, \uparrow,-}}{\eta_{-,\downarrow,-}}
-\frac{\eta_{+, \uparrow,-}}{\eta_{+,\downarrow,-}}\right]-\nonumber \\
(1-D) \left[\frac{\eta_{+,
\uparrow,+}}{\eta_{+,\downarrow,+}} -\frac{\eta_{-,
\uparrow,-}}{\eta_{-,\downarrow,-}}\right]
\left[\frac{\eta_{-, \uparrow,+}}{\eta_{-,\downarrow,+}}
-\frac{\eta_{+, \uparrow,-}}{\eta_{+,\downarrow,-}}\right].
\label{energyA-up-down}
\end{eqnarray}
The evident expression for $F_{\uparrow \downarrow}$ can be obtained by substituting Eq. (\ref{ratup-down2}) into
Eq. (\ref{energyA-up-down}), which yields,
\begin{equation}
\hspace{-2mm} F_{\uparrow \downarrow}=-\frac{16 B^2 e^{-i\theta}}{M^2_-}\Big\{\alpha^2 k^2-
\left(E^2+ \alpha^2 k^2\right)D\sin^2\frac{\theta}{2}\Big\}.
\label{FAup-down}
\end{equation}
Note that the expression for $F_{\downarrow \uparrow}(k)$ can be obtained from Eq. (\ref{energyA-up-down}) by replacing
$\alpha \to - \alpha$, $h \to - h$, $\theta \to - \theta$, and $M_- \to M_+$.
The main feature of the Andreev bound state energy in the anomalous particle-particle channel is that it survives
only in the presence of the in-plane magnetic field ${\bf B}$ and the spin-orbit interaction $\alpha$.
$E'_{\uparrow \downarrow}$ and $E'_{\downarrow \uparrow}$ vanish in the absence of one of the factors either
${\bf B}$ or $\alpha$, and they depend on the angle $\theta$ between the in-plane magnetic field orientation
and the junction, contributing to the fractional magneto-Josephson effect.

Andreev bound state energies and Josephson current, corresponding to different tunneling
channels, demonstrate completely different oscillation. The conditions
$F_{\uparrow \downarrow}^{\ast}(E_{\uparrow \downarrow})=0$ and $F_{\downarrow \uparrow}^{\ast}(E_{\downarrow \uparrow})=0$
with Eq. (\ref{Fup-down}) for $F_{\uparrow \downarrow}^{\ast}(E)$ provide  contributions to the Andreev bound state
energy in the particle-hole channel, which oscillates fractionally with the order parameters' phases difference
$\varphi$. Additional contributions to the energy come from the conditions
$F_{\uparrow \uparrow}^{\ast} (E_{\uparrow \uparrow})=0$ and $F_{\downarrow \downarrow}^{\ast} (E_{\downarrow \downarrow})=0$ with
$F_{\uparrow \uparrow}^{\ast}$ given by Eq. (\ref{Fup-up}),  which arise only in the presence of an in-plane
magnetic field ${\bf B}$ and oscillate not only with $\varphi$ but also with $\theta$. Contribution to
the magneto-Josephson effect gives apart from the anomalous particle-hole channel also the anomalous
particle-particle channel under the conditions $F_{\uparrow \downarrow}(E'_{\uparrow \downarrow})=0$ and
$F_{\downarrow \uparrow}(E'_{\downarrow \uparrow})=0$, where the evident expression for  $F_{\uparrow \downarrow}$
is given by Eq. (\ref{FAup-down}). Furthermore, the contribution coming from the anomalous
particle-particle channel vanishes not only at ${\bf B}=0$ but also in the
absence of the spin-orbit interaction, $\alpha =0$.
\begin{figure}[t]
\resizebox{.48\textwidth}{!}{%
\includegraphics[width=1cm]{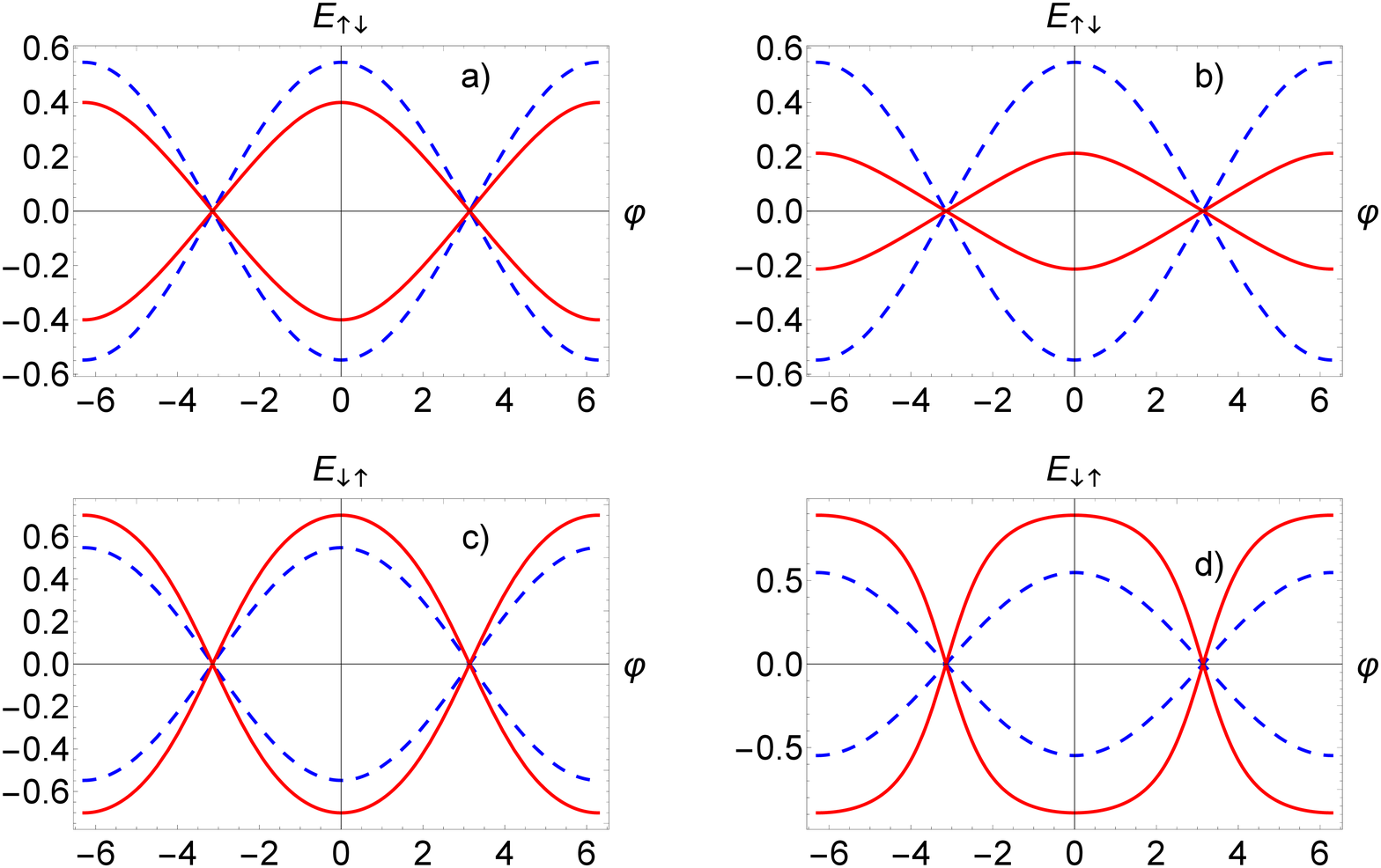}}
\caption {Andreev bound state energy $E_{\uparrow \downarrow}$ in the particle-hole channel for ${\bf B}=h=0$, $D=0.3$ and
(a) $\alpha=0.2$, (b) $\alpha=0.5$ and the energy  $E_{\downarrow \uparrow}$ under the same conditions for (c) $\alpha=0.2$
and (d) $\alpha=0.5$. Note that all energy parameters $E_{\uparrow \downarrow}$, $E_{\downarrow \uparrow}$, $B$, and $h$ in the figures
are given dimensionless in the scale of $\Delta$. $\alpha$ is also dimensionless in the scale of $v_F$.}
\label{B0-alpha}
\end{figure}
The total Andreev bound
state energy $E$ is obtained by finding overlap energies for each channel from the equations
$F_{\uparrow \downarrow}^{\ast}(E_{\uparrow \downarrow})=0$, $F_{\downarrow \uparrow}^{\ast}(E_{\downarrow \uparrow})=0$,
$F_{\uparrow \uparrow}^{\ast}(E_{\uparrow \uparrow})=0$, $F_{\downarrow \downarrow}^{\ast}(E_{\downarrow \downarrow})=0$, and
$F_{\uparrow \downarrow}(E'_{\uparrow \downarrow})=0$, $F_{\downarrow \uparrow}(E'_{\downarrow \uparrow})=0$,
and  summing up of all the coupling energies $E(\varphi, \theta)=E_{\uparrow \downarrow}(\varphi)+E_{\downarrow \uparrow}(\varphi)+
E_{\uparrow \uparrow}(\varphi, \theta) + E_{\downarrow \downarrow}(\varphi, \theta) + E'_{\uparrow \downarrow}(\theta) +
E'_{\downarrow \uparrow}(\theta)$ in each reflection channel. Below we calculate the Andreev bound state energies
for several asymptotic cases.
\begin{figure}[t]
\resizebox{.48\textwidth}{!}{%
\includegraphics[width=1cm]{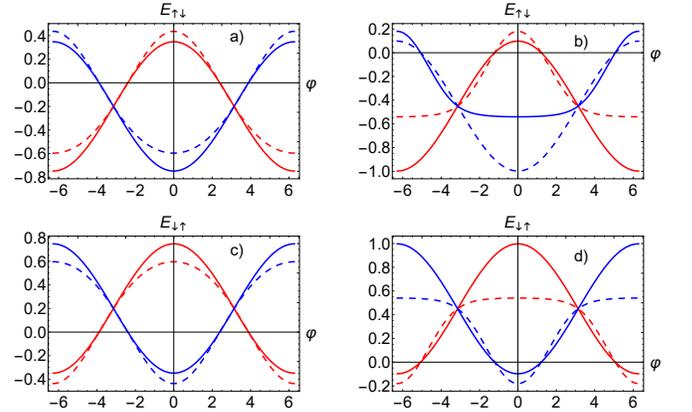}}
\caption {Andreev bound state energy $E_{\uparrow \downarrow}$ in the particle-hole channel under the conditions
${\bf B}=\alpha =0$, $D=0.3$,  for (a) $h=0.2$, (b) $h=0.45$, and  the energy $E_{\downarrow \uparrow}$ under
the same conditions, for (c) $h=0.2$, (d) $h=0.45$.}
\label{B0-h}
\end{figure}

\subsection{Andreev bound state energy in the absence of in-plane magnetic field, ${\bf B}=0$.}
Contribution to the Andreev bound state energy in the absence of {\it in-plane} magnetic field ${\bf B}=0$
comes only from the particle-hole scattering channel, determined by scattering amplitude
Eq. (\ref{ratup-down}), and all other channels vanish under this condition. The evident
expression for the bound state energy in the particle-hole channel is obtained from
the equation $F_{\uparrow \downarrow}^{\ast}(E_{\uparrow \downarrow})=0$, where $F_{\uparrow \downarrow}^{\ast}$ is given by
Eq. (\ref{Fup-down}). The general expression for $E_{\uparrow \downarrow}$ when all the external parameters
take non-zero values, $\alpha \neq 0$, ${\bf B} \neq 0$ and $h \neq 0$, can be obtained from the expression
(\ref{AEup-down}) in Appendix. By putting ${\bf B}=0$  in this equation and replacing $k^2$ according
to  Eq. (\ref{k2}) in Appendix one gets the following equation after routine calculations,
\begin{widetext}
\begin{eqnarray}
&&\left[(E_{\uparrow \downarrow}+h)^2\left(1-D \cos^2\frac{\varphi}{2}\right)-\left(\frac{v_F-\alpha}{v_F+\alpha}\right) (E^2_{\uparrow \downarrow}
-h^2 -\Delta^2)D \cos^2 \frac{\varphi}{2}\right]^2 +\nonumber\\
&&\frac{4D\cos^2\frac{\varphi}{2}}{(v_F +\alpha)^2} \left\{ (E_{\uparrow \downarrow} +h)^2 \left(1-D \cos^2\frac{\varphi}{2}\right)
\left[(E_{\uparrow \downarrow}v_F+\alpha h)^2 - \Delta^2 v_F^2\right] -
(v_F -\alpha)^2 h^2 \Delta^2 D \cos^2\frac{\varphi}{2}\right\}=0.
\label{E-B=0}
\end{eqnarray}
\end{widetext}
This equation is fourth order in $E_{\uparrow \downarrow}$ equation, and it can be in principle  solved analytically.
\begin{figure}[t]
\resizebox{.48\textwidth}{!}{%
\includegraphics[width=1cm]{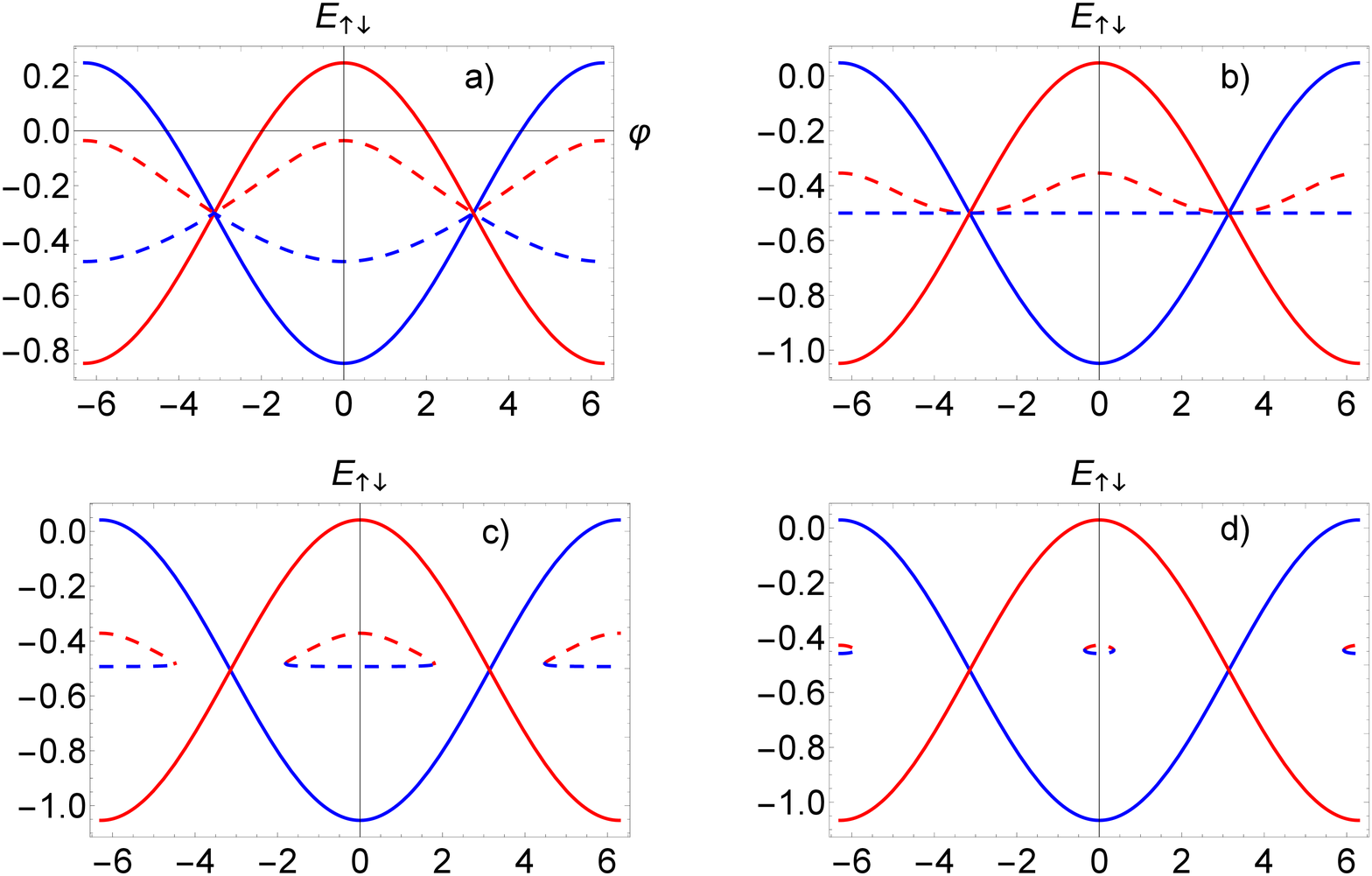}}
\caption {Andreev bound state energy $E_{\uparrow \downarrow}$ in the particle-hole channel for ${\bf B}=0$, $D=0.3$ and
$\alpha=0.4$ (a) $h=0.3$, (b) $h=0.5001$, (c) $h=0.506$, (d) $h=0.5185$.}
\label{B0-alpha-h}
\end{figure}
Equation (\ref{E-B=0}) yields exact analytical solutions for $E_{\uparrow \downarrow}$ in several asymptotic cases.

This equation is further simplified for \underline{\emph{ h=0, $\alpha$ =0, ({\bf B}=0)}}, yielding
\begin{equation}
E_{\pm} \equiv E_{\uparrow \downarrow}= \pm E_0 \equiv \pm |\Delta| \sqrt{D} \cos \frac{\varphi}{2},
\label{ksy}
\end{equation}
which reproduces the well-known result \cite{ksy04} for the Andreev bound state energy of $JJ$
with $p$-wave superconductors in the absence of magnetic field and spin-orbit interactions.
This expression provides the energy spectrum of quasi-electron and quasi-hole excitations,
symmetrically  located around the Fermi level in the gap.

\underline{\emph{In the case of $h=0$ and $\alpha \neq 0$ (${\bf B}=0$)}}, Rashba spin-orbit interaction
splits both quasi-electron and quasi-hole spectra, and  Eq. (\ref{E-B=0}) yields four solutions for the bound
state energy,
\begin{equation}
E=E_{\uparrow \downarrow}= s\frac{\sqrt{D} |\Delta| \frac{(v_F \pm \alpha)}{(v_F+\alpha)} \cos \frac{\varphi}{2}}
{\sqrt{1- D \cos^2\frac{\varphi}{2}\frac{(v_F+\alpha)^2 - (v_F \pm \alpha)^2}{(v_F +\alpha)^2}}},
\label{E-B=01}
\end{equation}
where $s = \pm$ assigns the electron and hole branches of the spectrum.
Two solutions of this expression coincides with Eq. (\ref{ksy}), and do not depend on the strength of Rashba
spin-orbit interaction. Nevertheless other two solutions depend on $\alpha$. The expression for Andreev's bound
state energy  $E_{\downarrow \uparrow}$, as mentioned above, is obtained by replacement of $\alpha \to -\alpha$ in
the expression (\ref{E-B=01}) written for  $E_{\uparrow \downarrow}$.    Figs.\ref{B0-alpha}a, b and
Figs.\ref{B0-alpha}c, d  depict the dependence of $E= E_{\uparrow \downarrow}$ and  $E_{\downarrow \uparrow}$ respectively
on $\varphi$ for two different values of $\alpha$ when $\alpha=0.2$ and $\alpha=0.5$. According to the figures,
two branches of Andreev's bound state energy, drawn by blue and dashed curves in  Figs.\ref{B0-alpha} do not
depend on $\alpha$. Nevertheless, other two branches of $E_{\uparrow \downarrow}$  (of $E_{\downarrow \uparrow}$) decrease
(increase) with increasing the strength of Rashba SOC. Note here that the parameters in all figures are given
in a dimensionless form as $E \to E/\Delta$, $k \to (kv_F)/\Delta$, $h \to h/\Delta$, ${\bf B} \to {\bf B}/\Delta$,
and $\alpha \to \alpha/v_F$. In this limiting case, the quasi-electron and quasi-hole spectra are again
symmetrically located around the Fermi level.
\begin{figure}[t]
\resizebox{.48\textwidth}{!}{%
\includegraphics[width=1cm]{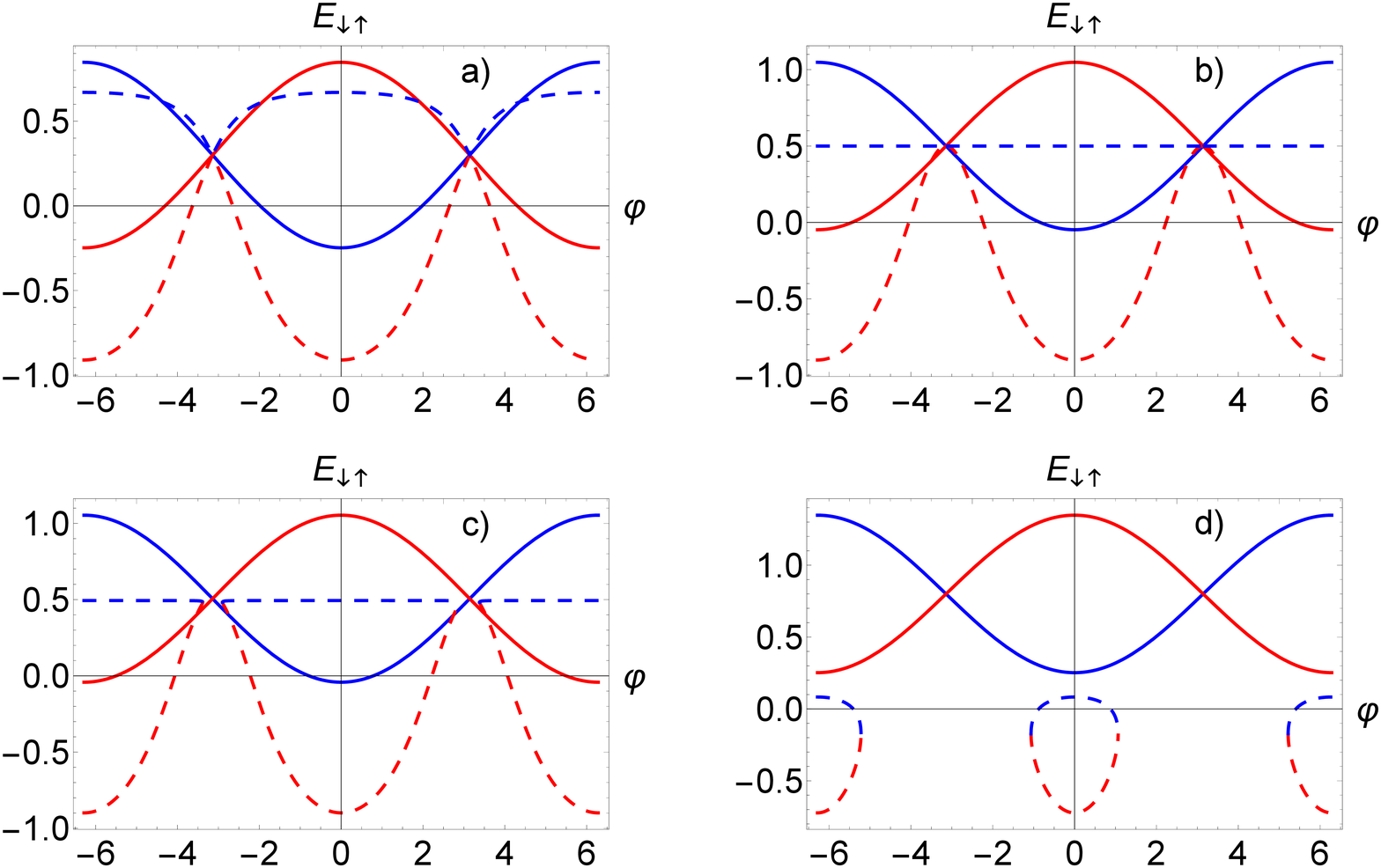}}
\caption {Andreev bound state energy $E_{\downarrow \uparrow}$ in the particle-hole channel for ${\bf B}=0$, $D=0.3$ and
 $\alpha=0.4$, (a) $h=0.3$, (b) $h=0.5001$, (c) $h=0.506$, (d) $h=0.8$.}
\label{B0-h-alpha}
\end{figure}

\underline{ \emph{In the case of $\alpha =0$ and $h \neq 0$ (${\bf B}=0$)}}
 Equation (\ref{E-B=0}) yields the following expression for the Andreev bound state energy in this limit,
\begin{eqnarray}
E_{\uparrow \downarrow}^+ &=& -h \left[1-2 D
\cos^2\frac{\varphi}{2}\right] \nonumber\\
&& + s \sqrt{D} \cos\frac{\varphi}{2}
\sqrt{\Delta^2-4h^2 + 4D h^2 \cos^2 \frac{\varphi}{2}} \nonumber\\
E_{\uparrow \downarrow}^- &=& -h + s \sqrt{D} \sqrt{\Delta^2-4h^2}
\cos\frac{\varphi}{2} \label{E-B=02}
\end{eqnarray}
Andreev bound states are split again due to Zeeman effect. The
dependence of  $E_{\uparrow \downarrow}^{\pm} \equiv E_{\uparrow
\downarrow}$ on $\varphi$ is depicted in Figs. \ref{B0-h} a, b for
two different values of $h$. Note that contribution to Andreev bound
state energy $E_{\downarrow \uparrow}$, found from the condition of
$F^{\ast}_{\downarrow \uparrow}=0$, can be obtained again by
replacing $h \to -h$, $\alpha \to - \alpha$, and $\theta \to -
\theta$ in Eqs. (\ref{ksy}), (\ref{E-B=01}), and (\ref{E-B=02}). The
dependence of $E_{\downarrow \uparrow}$ on $\varphi$ for different
values of $h$ is depicted in Figs. \ref{B0-h} c and d for
completeness. Two branches of solution (\ref{E-B=02}) differ from
those given by Eq. (\ref{ksy}) by shifting only the particle and
hole pairs $E_{\uparrow \downarrow}$ ($E_{\downarrow \uparrow}$) to
the value of $-h$ ( $+h$), without changing their oscillation
characteristics (see, Figs. \ref{B0-h} a, b and c, d). As it is seen
clearly from Figs. \ref{B0-h} a, b (Figs. \ref{B0-h} c, d)  the
magnetic field reduces considerably the amplitude of the fractional
oscillation for other two solutions of $E_{\uparrow \downarrow}$
($E_{\downarrow \uparrow}$), at the same time shifts down (up)
asymmetrically the quasi-particle and quasi-hole spectra. One of the
quasi-hole (quasi-particle) branch of $E_{\uparrow \downarrow}$
($E_{\downarrow \uparrow}$) is pushed off from the gap at higher
magnetic field when $h>h_c=0.46 \Delta$.
\begin{figure}[t]
\resizebox{.48\textwidth}{!}{%
\includegraphics[width=1cm]{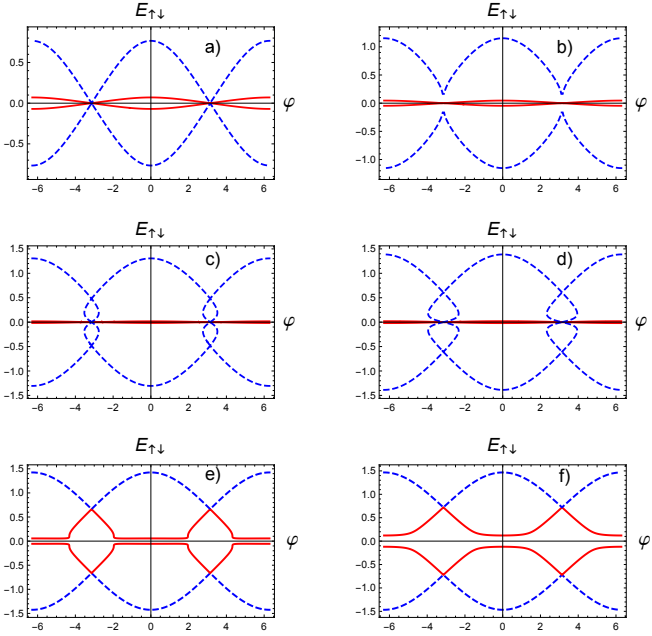}}
\caption {Andreev bound state energy $E_{\uparrow \downarrow}$ in the particle-hole channel for $h=0$, $D=0.3$ and
$\alpha=0.3$ (a) $B=0.5$, (b) $B=0.83$, (c) $B=0.96$, (d) $B=1.03$, (e) $B=1.063$, and (f) $B=1.1$.}
\label{h0-alpha-B}
\end{figure}

\underline{\emph{The general case for ${\bf B}=0$, but $\alpha \neq 0$ and $h \neq 0$}} is calculated
numerically according to Eq. (\ref{E-B=0}) writing this equation in the dimensionless parameters such
as $\tilde{E}= E/\Delta$, $\tilde{k}=(kv_F)/\Delta$, $\tilde{h}=h/\Delta$,
$\tilde{{\bf B}}={\bf B}/\Delta$, and $\tilde{\alpha}=\alpha/v_F$.  Fig. \ref{B0-alpha-h} shows the dependence
of $E_{\uparrow \downarrow}$ on $\varphi$ for $D=0.3$ and $\tilde{\alpha}=0.4$ with different values of
$\tilde{h}$, $\tilde{h}=0.3; 0.5001; 0.506$; and $0.5185$ (the parameters in all figures are given without tilde).
One of the quasi-electron and quasi-hole pair of the
spectrum, depicted by solid (blue and red) lines in Fig. \ref{B0-alpha-h}, shifts down with increasing the magnetic
field $h$ without changing the form and amplitude.  The amplitude of the other quasi-electron and quasi-hole branch
of $E_{\uparrow \downarrow}$ (drawn by dashed blue and red curves) decreases, and  the form of the curves is deformed
with increasing the magnetic field $h$. At $\tilde{h} > \tilde{h}_g=0.5001$ a forbidden gap appears in the spectrum,
i.e. as it is seen in Fig. \ref{B0-alpha-h}c the quasi-electron and quasi-hole states disappear for some values of
the order parameter phase difference $\varphi$. The quasi-particle and quasi-hole states, shown by dashed (blue and red)
curves in Fig. \ref{B0-alpha-h} vanishes by further increasing of the magnetic field at $\tilde{h} > \tilde{h}_c=0.51921$.
\begin{figure}[t]
\resizebox{0.48\textwidth}{!}{%
\includegraphics[width=1cm]{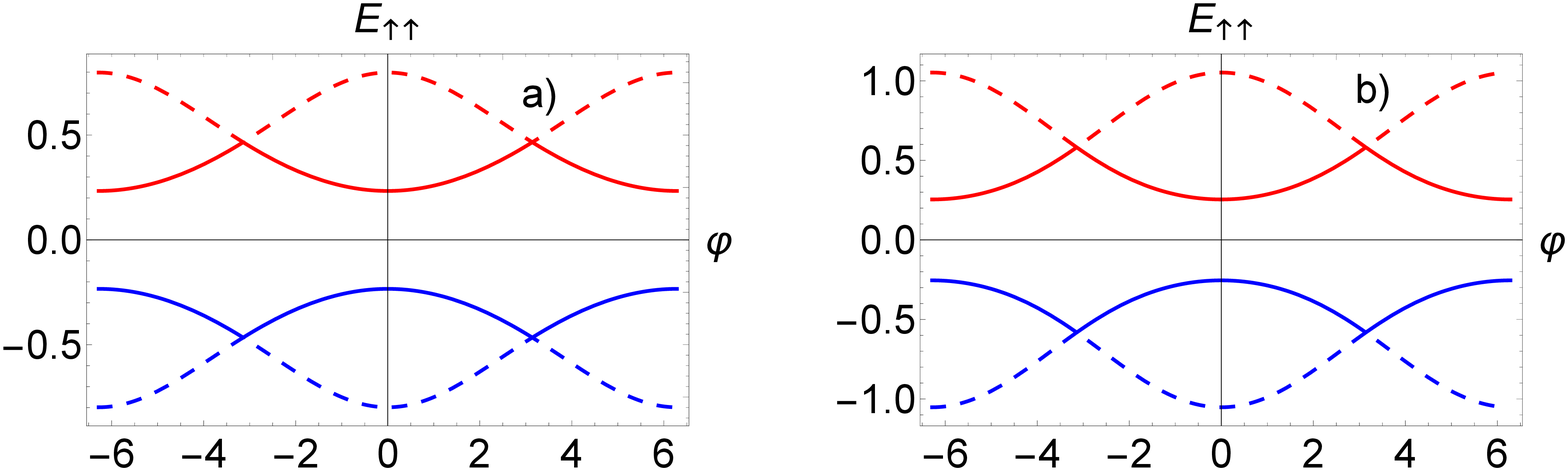}}
\caption{Andreev bound state energy $E_{\uparrow \uparrow}$ in the anomalous particle-hole channel for $h=0$,
$D=0.3$ and  $\alpha=0.2$, (a) $B=0.2$, (b) $B=0.5$.}
\label{Up-alpha-B}
\end{figure}
For completeness, $E_{\downarrow \uparrow}$ vs. $\varphi$ dependence is calculated also for
$h=0.3; 0.5001; 0.506$; and $0.5185$ under the condition of  ${\bf B}=0$, $\alpha=0.4$
and $D=0.3$, which is depicted in  Fig. \ref{B0-h-alpha}. As it is expected, $E_{\downarrow \uparrow}$ behaves
like  $E_{\uparrow \downarrow}$, i.e. the magnetic field shifts up one of the quasi-partice and quasi-hole pair,
drawn by solid blue and red curves in Fig. \ref{B0-h-alpha} without changing the amplitude and form. The other pair,
presented by dashed blue and red curves  Fig. \ref{B0-h-alpha} deforms and amplitude decreases with increasing the
magnetic field $h$. For $h>0.5001$ a forbidden gap is opened (see, Fig. \ref{B0-h-alpha} b) in the spectrum.
This branch (dashed curves in Fig. \ref{B0-h-alpha} c, d) squeezes and disappears for $h>h_c=0.91124$.

\subsection{Andreev bound state energy in the presence of {\bf \it in-plane} magnetic field ${\bf B} \neq 0$}
In the presence of the in-plane magnetic field ${\bf B}$ all three channels described by
Eqs. (\ref{ratup-down})- (\ref{up-up}) give contributions to Andreev bound state energy $E$.
The  expressions for general dependencies of $E_{\uparrow \downarrow}$, $E_{\uparrow \uparrow}$ and
$E'_{\uparrow \downarrow}$  on $\alpha,~ {\bf B}, ~ h$ can be obtained from the Equations (\ref{AEup-down}),
(\ref{AEup-up2}) and (\ref{FAup-down1}) presented in Appendix after replacement of $k^2$ by $E$
according to Eq. (\ref{k2}). These equations can be solved analytically for energy in several
asymptotic cases. Note that main contribution to Andreev bound state energy still gives the conventional
particle-hole channel.
\begin{figure}[t]
\resizebox{.48\textwidth}{!}{%
\includegraphics[width=1cm]{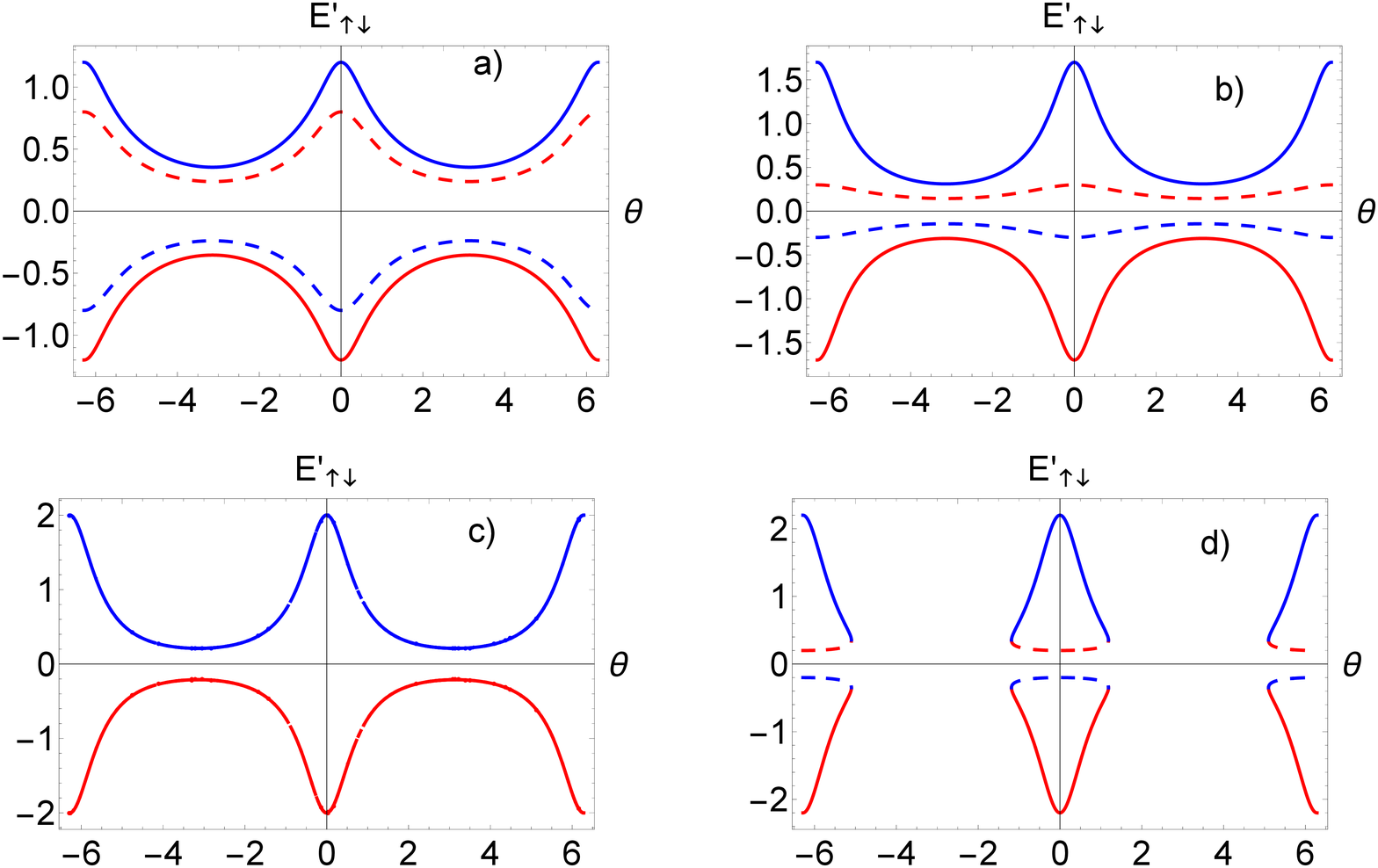}}
\caption {Andreev bound state energy $E'_{\uparrow \downarrow}$ in the anomalous particle-particle channel for $h=0$, $D=0.3$ and
 $\alpha=0.2$, (a) $B=0.2$, (b) $B=0.7$, (c) $B=1.0$, and (d) $B=1.2$. }
\label{F3-h0}
\end{figure}

\underline{\emph{The case of $\alpha$ =0, ${\bf B} \neq 0$,and  $h \neq 0$.}}
In this limiting case an interference between SOC-induced effective magnetic field and $h$ vanishes,
and hence the energy spectrum depends on the modulus of total magnetic field according to Eq. (\ref{E4})
as $H= \sqrt{{\bf B}^2+ h^2}$. The expression (\ref{AEup-down}) for $E_{\uparrow \downarrow}$ is
strongly simplified in this limiting case, and substitution of $k^2$ from Eq. (\ref{E4}) into
this expression yields,
\begin{equation}
E_{\uparrow \downarrow}^{\pm}= \pm \sqrt{B^2+h^2} +s \Delta \sqrt{D}
\cos \frac{\varphi}{2}, \label{Eup-down1}
\end{equation}
where $s = \pm$.
\begin{figure}[t]
\resizebox{.48\textwidth}{!}{%
\includegraphics[width=1cm]{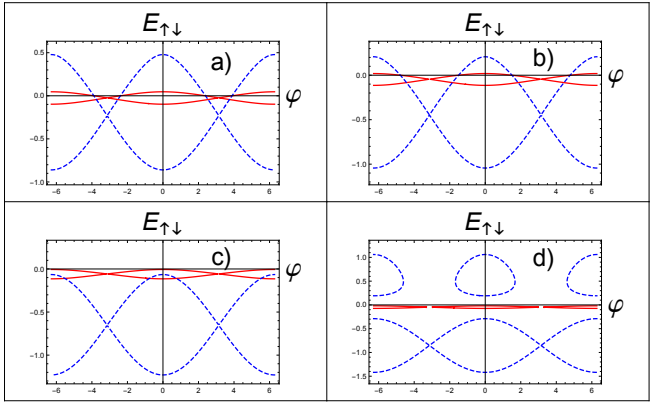}}
\caption {Andreev bound state energy $E_{\uparrow \downarrow}$ in the  particle-hole channel for $D=0.3$,
 $\alpha=0.3$, and $B=0.4$ (a) $h=0.2$, (b) $h=0.4$, (c) $h=0.6$, and (d) $h=0.8$. }
\label{Fup-downBah}
\end{figure}
The Andreev bound state energy in the anomalous particle-hole channel $E_{\uparrow \uparrow}$  can be found in
this limiting case from the general expression given by Eq. (\ref{AEup-up3}) yielding,
\begin{equation}
E_{\uparrow \uparrow}^{\pm}= \pm \sqrt{B^2+h^2} +s \Delta \sqrt{D}
\cos \frac{\varphi-\theta}{2}. \label{Eup-up1}
\end{equation}
This expression differs from that given by (\ref{Eup-down1}) for $E_{\uparrow \downarrow}$  by dependence
of cosine function not only on $\varphi$ but also on $\theta$.

Contribution to Andreev bound state energy from the third anomalous particle-particle channel vanishes,
as is seen from Eq. (\ref{FAup-down2}), in this limiting case. So, one can state that the third channel
survives and gives a contribution to the bound state energy only in the presence of SOC ($\alpha \neq 0$)
and in-plane magnetic field (${\bf B} \neq 0$) in the system. An absence at least one of these factors
destroys this channel.
\begin{figure}[t]
\resizebox{.48\textwidth}{!}{%
\includegraphics[width=1cm]{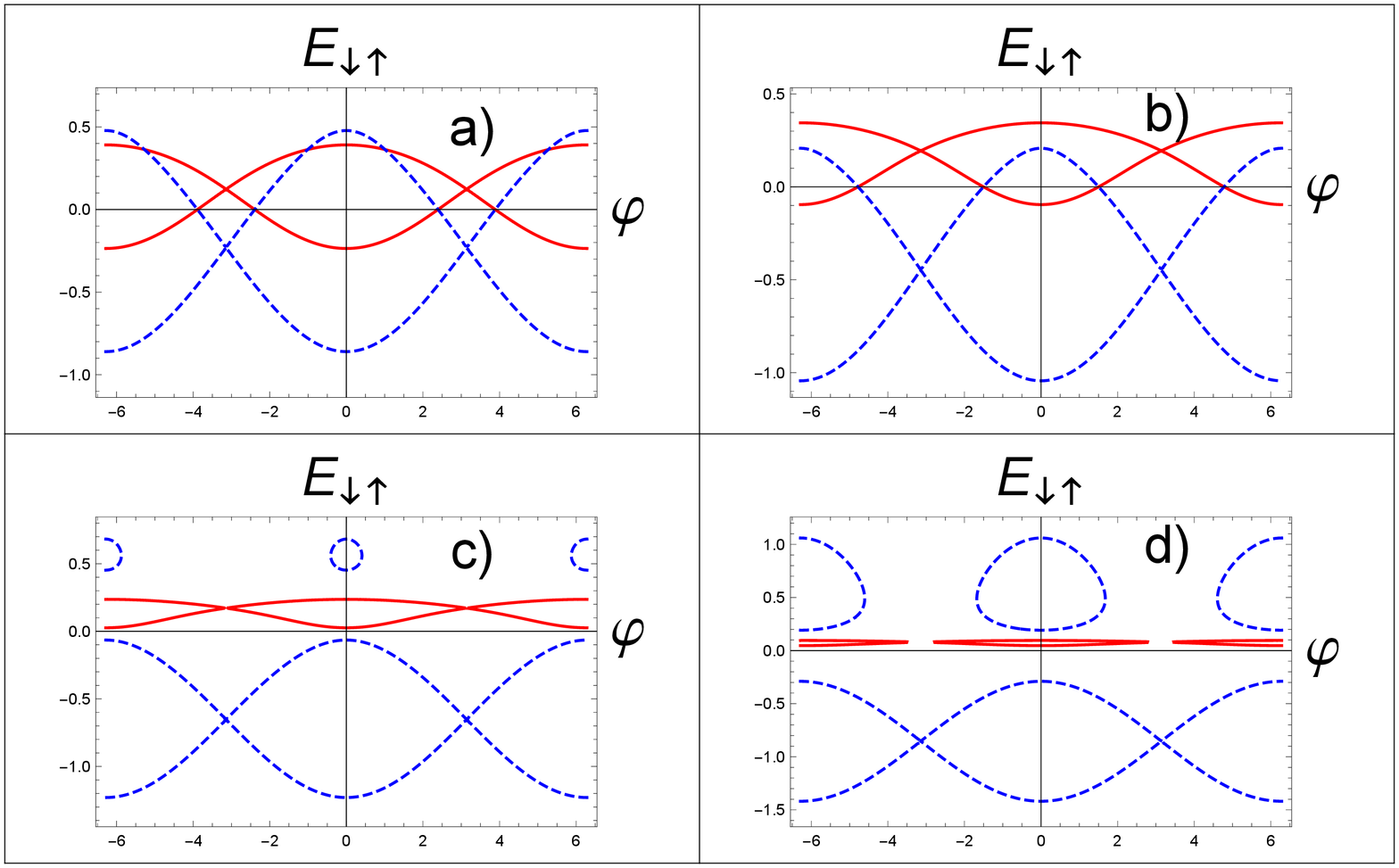}}
\caption {Andreev bound state energy $E_{\downarrow \uparrow}$ in the particle-hole channel for $D=0.3$,
 $\alpha=0.3$, and $B=0.4$ (a) $h=0.2$, (b) $h=0.4$, (c) $h=0.6$, and (d) $h=0.8$. }
\label{Fdown-upBah}
\end{figure}
The total Andreev bound state energy in this limiting case  contains (\ref{Eup-down1}) and (\ref{Eup-up1}),
and also the energies $E_{\downarrow \uparrow}$ and $E_{\downarrow \downarrow}$, obtained from (\ref{Eup-down1})
and (\ref{Eup-up1}) by replacements $\alpha \to - \alpha$, $h \to -h$, $\theta \to - \theta$,
\begin{eqnarray}
&&E=  \pm 4 \sqrt{B^2+h^2} + s \Delta \sqrt{D} \Bigg\{2  \cos \frac{\varphi}{2} +\nonumber\\
&& \cos \frac{\varphi-\theta}{2} +  \cos \frac{\varphi+\theta}{2} \Bigg\}.
\label{E1}
\end{eqnarray}

\underline{\emph{The case of $h=0$, $\alpha \neq 0$ and ${\bf B} \neq 0$}}.
The expression  $F_{\uparrow \downarrow}^{\ast}=0$ can be simplified for $h=0$ and $\alpha \neq 0$, ${\bf B} \neq 0$.
Routine calculations yield the following expression to determine $E_{\uparrow \downarrow}$,
\begin{eqnarray}
&& E^2_{\uparrow \downarrow} \Big\{k^2\left[v_F \alpha (B^2 +\Delta^2 -E^2_{\uparrow \downarrow}) -v_F^2B^2\right]-\nonumber\\
&&\frac{v_F \alpha}{(v_F +\alpha)^2}(E^2_{\uparrow \downarrow}-B^2-\Delta^2)^2+\Delta^2 B^2\Big\}-\nonumber\\
&& D \Delta^2 \cos^2 \frac{\varphi}{2}\Big\{k^2\alpha \left[v_F(\Delta^2-E^2_{\uparrow \downarrow})-B^2 (v_F- \alpha)\right]-\nonumber\\
&& \frac{v_F \alpha}{(v_F+\alpha)^2}\left[(E^2_{\uparrow \downarrow}+\Delta^2-B^2)^2-4E^2_{\uparrow \downarrow}\Delta^2\right]+\nonumber\\
&& \left(\frac{v_F-\alpha}{v_F+\alpha} \right)^2E^2_{\uparrow \downarrow}B^2\Big\}=0
\label{h=0}
\end{eqnarray}
One can put the expression for $k^2_{\pm}$ from (\ref{k2}) and solve numerically this equation for $E_{\uparrow \downarrow}$.
Fig. \ref{h0-alpha-B} shows the dependence of $E_{\uparrow \downarrow}$ on different values of the in-plane magnetic field
$B$ for particular value of $\alpha=0.3$ and $h=0$. One quasi-particle and quasi-hole pair in the spectrum, depicted
in blue (dashed lines) is enlarged and is partially pushed off from the gap with increasing the in-plane
magnetic field $B$ at $B>B_c \approx 0.55$. On the other hand the pair, depicted in red in Fig. \ref{h0-alpha-B},
is narrowed with increasing $B$, and the gap is opened in the spectrum at $B>B_c \approx 0.95$. Further increase
in $B$ makes this branch of the spectrum again regular at $B>1.063$.
\begin{figure}[t]
\resizebox{.48\textwidth}{!}{%
\includegraphics[width=1cm]{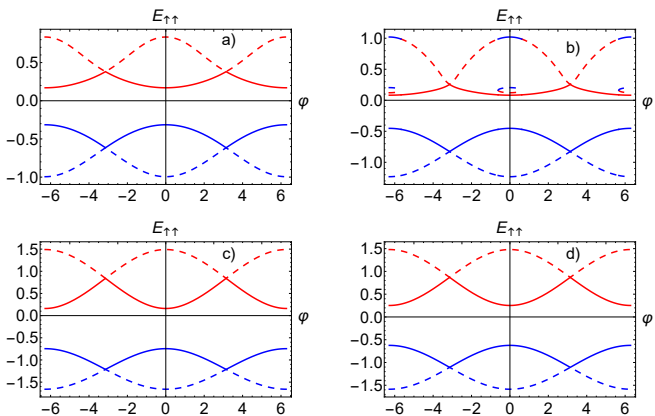}}
\caption {Andreev bound state energy $E_{\uparrow \uparrow}$ in the particle-particle channel for $D=0.3$,
 $\alpha=0.2$, (a) $B=0.3$, $h=0.2$, (b) $B=0.3$, $h=0.5$, (c) $B=0.3$, $h=0.98$, and (d) $B=0.8$, $h=0.58$.}
\label{FUpUpBah}
\end{figure}

The dependence of $E_{\uparrow \uparrow}$ on $\varphi$ in the anomalous particle-hole channel for non-zero values of
the external parameters ${\bf B}$ and $\alpha$ but for $h=0$
is depicted in Fig. \ref{Up-alpha-B} for $\alpha=0.2$, $h=0$, $D=0.3$ (a) $B=0.2$ and (b) $B=0.5$. The quasi-electron
(quasi-hole) dispersion at $h=0$ is shifted to higher (lower) values with increasing ${\bf B}$ and/or $\alpha$
without changing the shape and symmetry of the energy spectrum.

The third anomalous particle-particle channel gives a contribution $E'$ to the Andreev bound state energy, hence
to fractional magneto-Josephson effect provided that both parameters ${\bf B}$ and $\alpha$ take non-zero values.
Contribution to $E'$ now is calculated according to Eqs. (\ref{FAup-down1}) and (\ref{k2}). The numerical calculations
for $E'$ dependence on $\theta$ for the case of $h=0$  when $B$ takes different values is presented in Fig. \ref{F3-h0}.
According to Fig. \ref{F3-h0} one of the quasi-particle and quasi-hole branch drawn by solid blue and red curves enlarges
with in-plane magnetic field ${\bf B}$, nevertheless the (particle-hole) symmetry is preserved for all curves. The other
branch of the spectrum drawn by dashed red and blue curves squees and disappears (see, Fig. \ref{F3-h0}c) when $B$ approachs
unity.  For higher values of $B$ the forbidden gap (shown in Fig. \ref{F3-h0}d) appears in the spectrum.
\begin{figure}[t]
\resizebox{.48\textwidth}{!}{%
\includegraphics[width=1cm]{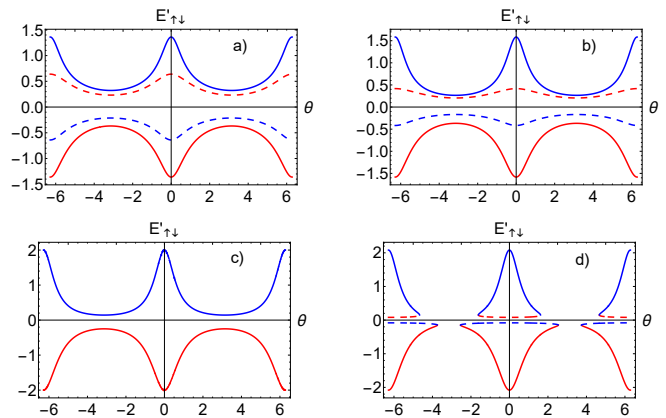}}
\caption {Andreev bound state energy $E'_{\uparrow \downarrow}$ in the anomalous particle-hole channel
for $D=0.3$,  $\alpha=0.2$, (a) $h=0.2$, $B=0.3$ (b) $h=0.5$, $B=0.3$ (c) $h=0.8$, $B=0.6$
(d) $h=0.9$, $B=0.6$.}
\label{F3-Bah}
\end{figure}

\underline{\emph{In the case of $h \neq 0$, $\alpha \neq 0$ and ${\bf B} \neq 0$}} out-of-plane magnetic field $h$
destroys a particle-hole symmetry in the spectrum. Dependence of $E_{\uparrow \downarrow}$ and $E_{\downarrow \uparrow}$ on
$\varphi$ in the particle-hole channel is depicted in Figs. \ref{Fup-downBah} and \ref{Fdown-upBah} for finite $B=0.4$
and different values of h, $h=0.2$, $h=0.4$, $h=0.6$, and $h=0.8$. The magnetic field seemly does not change the
amplitude and structure one of the quasi-particle and quasi-hole energy pair, drawn by blue and dashed curves
in  Figs. \ref{Fup-downBah} and \ref{Fdown-upBah}. These bound state energies are shifted along the energy axis only.
Instead, the magnetic fields strongly change other quasi-particle and quasi-hole pair, presented by red and solid
curves in  Figs. \ref{Fup-downBah} and \ref{Fdown-upBah}. This pair of the bound state energy is reduced in amplitude
with increasing the magnetic field. At $h>h_c=0.623$ a forbidden gap appears in the spectrum, and $h_c$ increases
with increasing $B$.

Numerical calculation of the Andreev bound state energy $E_{\uparrow \uparrow}$ in the anomalous particle-hole
channel is shown in Fig.\ref{FUpUpBah}. The dashed (red and blue) curves, corresponding to spin-up branches
of the bound energy, move away each other with increasing the magnetic fields. Instead the solid (blue and red)
curves, corresponding to the spin-down branch's of the spectrum, is slightly narrowed with increasing the magnetic field.

Andreev bound state energy $E'_{\uparrow \downarrow}$ in anomalous particle-particle channel is calculated for non-zero
values of $\alpha$ and magnetic fields ${\bf B}$ and $h$, the result of which is presented in Fig. \ref{F3-Bah}.
The solid (blue and red) curves, corresponding to spin-up branch of the spectrum in the figure inclreases in amplitude
with increasing the magnetic fields up to values $\sqrt{h^2+B^2}=1$. Instead the dashed (blue and red) curves,
corresponding  to spin-down branches of the spectrum, are narrowed and disappear at $\sqrt{h^2+B^2}=1$.
A gap is opened in the spectrum with further increase in the magnetic fields.

\section{Equilibrium Josephson current and spin current}
\label{sec4}
Josephson current carried by a quasi-particle state $a$ at zero temperature is
\begin{equation}
J_a=\frac{2e}{\hbar}\frac{\partial E_a}{\partial \varphi}
\label{current1}
\end{equation}
The current flowing thought the quasi-particle and quasi-hole states $J_{\pm}$ in
the simplest case of  ${\bf B} = h =0$ but $\alpha \neq 0$  can be obtained from the tunneling
energy given by Eq. (\ref{E-B=01}). For $J_+$, when the Andreev bound state energy becomes
\begin{equation}
E_{\uparrow \downarrow}^+=s\sqrt{D}\Delta \cos \frac{\varphi}{2}
\label{Eksy1}
\end{equation}
with $s=\pm$ assigning the quasi-particle and quasi-hole pair, one gets,
\begin{equation}
J_{s}^+=-s \frac{e \Delta \sqrt{D}}{\hbar} \sin \frac{\varphi}{2}.
\label{ksy1}
\end{equation}
For other particle-hole pair of the bound state energy  Eq. (\ref{E-B=01})
\begin{equation}
E_{\uparrow \downarrow}^-=s\sqrt{D} \Delta \frac{\frac{v_F-\alpha}{v_F+\alpha}}
{\sqrt{1-D\frac{4v_F \alpha}{(v_F+\alpha)^2}\cos^2\frac{\varphi}{2}}}\cos \frac{\varphi}{2}
\label{Eksy2}
\end{equation}
Josephson  energy $J_s^-$ reads as
\begin{equation}
J_s^-=-s\frac{e\sqrt{D}\Delta}{\hbar}\frac{\frac{v_F-\alpha}{v_F +\alpha}}{\left[1-D\frac{4v_F\alpha}{(v_F+\alpha)^2}
\cos^2 \frac{\varphi}{2}\right]^{3/2}}\sin \frac{\varphi}{2}.
\label{ksy2}
\end{equation}

In thermodynamic equilibrium at temperature $T$ the total contribution of the Andreev
bound states to the Josephson current can be calculated according to the expression
\begin{equation}
J=\frac{2e}{\hbar}\sum_{n=\pm} \frac{\partial E_{\uparrow \downarrow}^n}{\partial \varphi}~f(E_{\uparrow \downarrow}^n)=
- \frac{2e}{\hbar}\sum_{n= \pm}\frac{\partial E_{\uparrow \downarrow}^n}{\partial \varphi}\tanh\left(\frac{E_{\uparrow \downarrow}^n}{2k_BT}\right),
\label{current}
\end{equation}
where the expression of $E_{\uparrow \downarrow}^{\pm}$ are presented by Eqs. (\ref{Eksy1}) and (\ref{Eksy2}). Taking into accout
the expressions for the energies one gets for Josephson current in the simplest case when $B=h=0$ and $\alpha \neq 0$
\begin{eqnarray}
\hspace{-5mm}&&J=\frac{e\sqrt{D} \Delta}{\hbar}\sin \frac{\varphi}{2} \Bigg\{\tanh \frac{\sqrt{D}\Delta \cos \frac{\varphi}{2}}{2k_BT} +\nonumber\\
\hspace{-7mm}&&\frac{\frac{v_F-\alpha}{v_F+\alpha}}{\left[1-\frac{4v_F \alpha D \cos^2\frac{\varphi}{2}}{(v_F+\alpha)^2} \right]^{3/2}}
\tanh \frac{\sqrt{D} \Delta \frac{v_F -\alpha}{v_F+\alpha} \cos \frac{\varphi}{2}}
{2k_BT \sqrt{1-\frac{4v_F\alpha D \cos^2\frac{\varphi}{2}}{(v_F+\alpha)^2}}}\Bigg\}.
\end{eqnarray}
Josephson current in this case will depend on the Rashba SOC coefficient $\alpha$, which can be
experimentally determined.

Josephson current in the case of $B=\alpha=0$ and $h \neq 0$ can be calculated by using the expressions \ref{E-B=02}
for Andreev bound state energies $E^{\pm}_{\uparrow \downarrow}$. In this case the magnetic field $h$ makes asymmetric
the bound energy. For simplicity we calculate  here the total Josephson current
\begin{equation}
 I(h, T)=I_{\uparrow \downarrow}+I_{\downarrow \uparrow}
\end{equation}
which correspon to bound state energies $E^{\pm}_{\uparrow \downarrow}$ and $E^{\pm}_{\downarrow \uparrow}$ respectively.
The routine calculations yield
\begin{widetext}
\begin{eqnarray}
I(h, T)=\frac{2e}{\hbar}\left[hD\sin \varphi +\frac{\sqrt{D}}{2}\sin \frac{\varphi}{2}\frac{{\mathbb D}^2 +
4h^2D\cos^2 \frac{\varphi}{2}}{{\mathbb D}} \right] \tanh \frac{-h\left(1-2D\cos^2\frac{\varphi}{2}\right)+
\sqrt{D} {\mathbb D}\cos \frac{\varphi}{2}}{2k_BT}-\nonumber\\
\frac{2e}{\hbar}\left[hD\sin \varphi - \frac{\sqrt{D}}{2}\sin \frac{\varphi}{2} \frac{{\mathbb D}^2 +
4h^2D\cos^2 \frac{\varphi}{2}}{{\mathbb D}} \right] \tanh \frac{h\left(1-2D\cos^2\frac{\varphi}{2}\right)+
\sqrt{D} {\mathbb D}\cos \frac{\varphi}{2}}{2k_BT}+\nonumber\\
\frac{e}{\hbar}\Delta \sqrt{D} \sin \frac{\varphi}{2}\left[\tanh \frac{-h +\Delta \sqrt{D} \cos \frac{\varphi}{2}}{2k_BT} +
\tanh \frac{h +\Delta \sqrt{D} \cos \frac{\varphi}{2}}{2k_BT} \right],
\label{current-h}
\end{eqnarray}
\end{widetext}
where  ${\mathbb D}=\sqrt{\Delta^2 -4h^2\left(1-D\cos^2 \frac{\varphi}{2}\right)}$.
In two limiting case this expression is simplified. At $D \to 0$ and $h>k_BT$ Eq. (\ref{current-h}) yields
\begin{equation}
I(h, T) =-\frac{4e}{\hbar} h D \sin \varphi.
\end{equation}
In the opposite case, when $D \to 0$  and $h< k_BT$ one gets,
\begin{equation}
I=\frac{e}{\hbar}D \frac{\Delta^2 -2h^2}{k_BT} \sin \varphi.
\end{equation}

The Josephson current is given as a partial derivative of the system's energy with
respect to the superconducting phase $\varphi$ as $J_Q=\frac{2e}{\hbar}\frac{\partial \langle H \rangle}{\partial \varphi}$,
where $H$ is the system's Hamiltonian.  In the case of topological insulator edges, the spin currents
arise as the exact duals of the Josephson current, $J_S =\frac{\partial \langle H \rangle}{\partial \theta}$.
We define $\theta$  as the angle between the wire and the Zeeman field, which is also exact dual to the
superconducting phase $\varphi$. Thereby, spin Josephson currents $j_S$ are equivalent to torques \cite{bb09}
(driven partly by the Majoranas) that the wire domains apply on the external magnets. Our calculations allow
us to find the spin current. Indeed, Andreev bound state energies in anomalous particle-hole channel
$E_{\uparrow \uparrow}(\varphi, \theta)$, $E_{\downarrow \downarrow}(\varphi, \theta)$  and in anomalous particle-particle
channel $E'_{\uparrow \downarrow}(\theta)$, $E'_{\downarrow \uparrow}(\theta)$  give contribution to the spin-Josephson
current, which oscillate with $4\pi$ periodicity.

\section{AC Josephson Effect}
\label{sec5}

\begin{figure}[t]
\includegraphics[width=4.7cm]{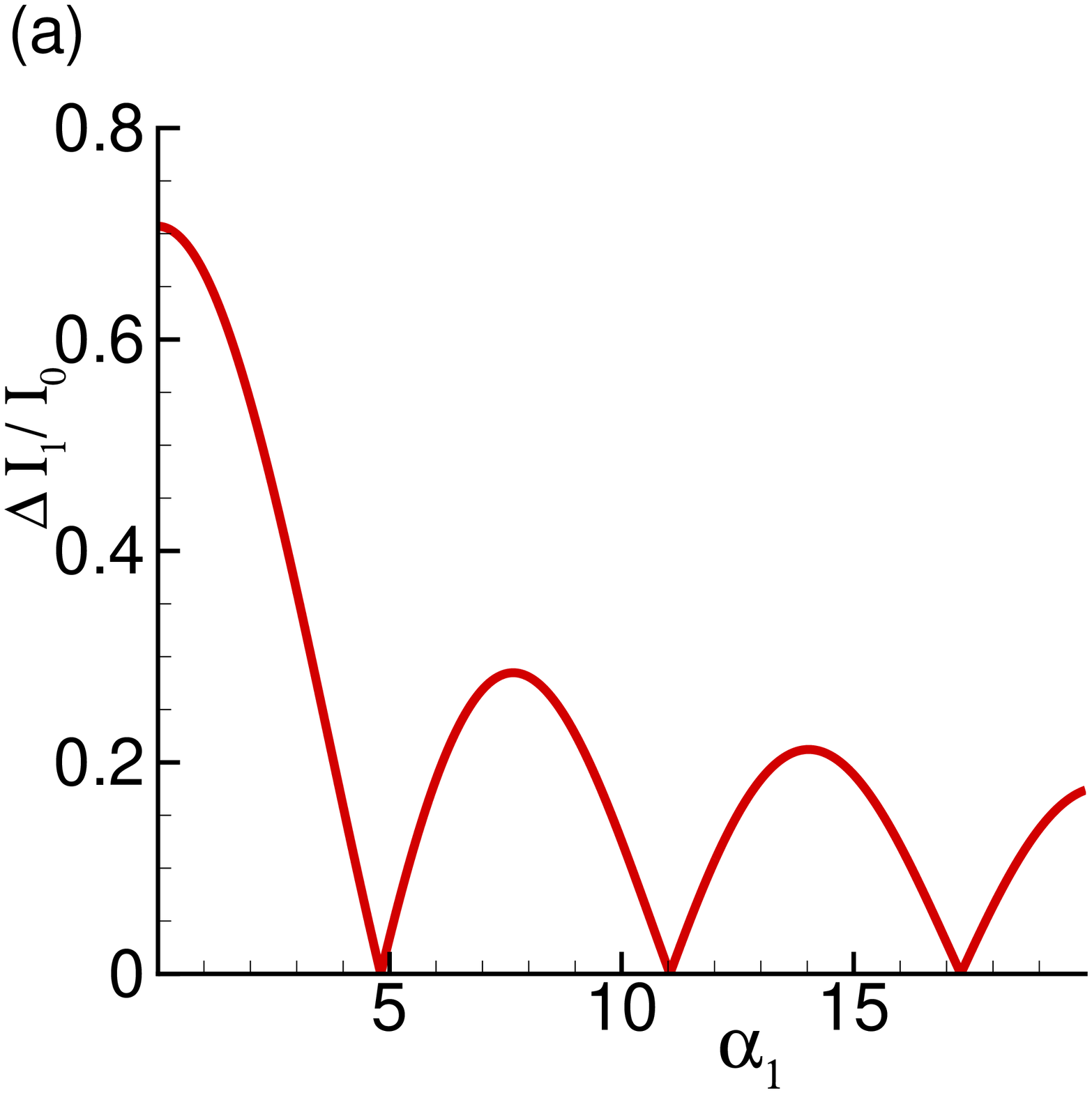}\includegraphics[width=4.7cm]{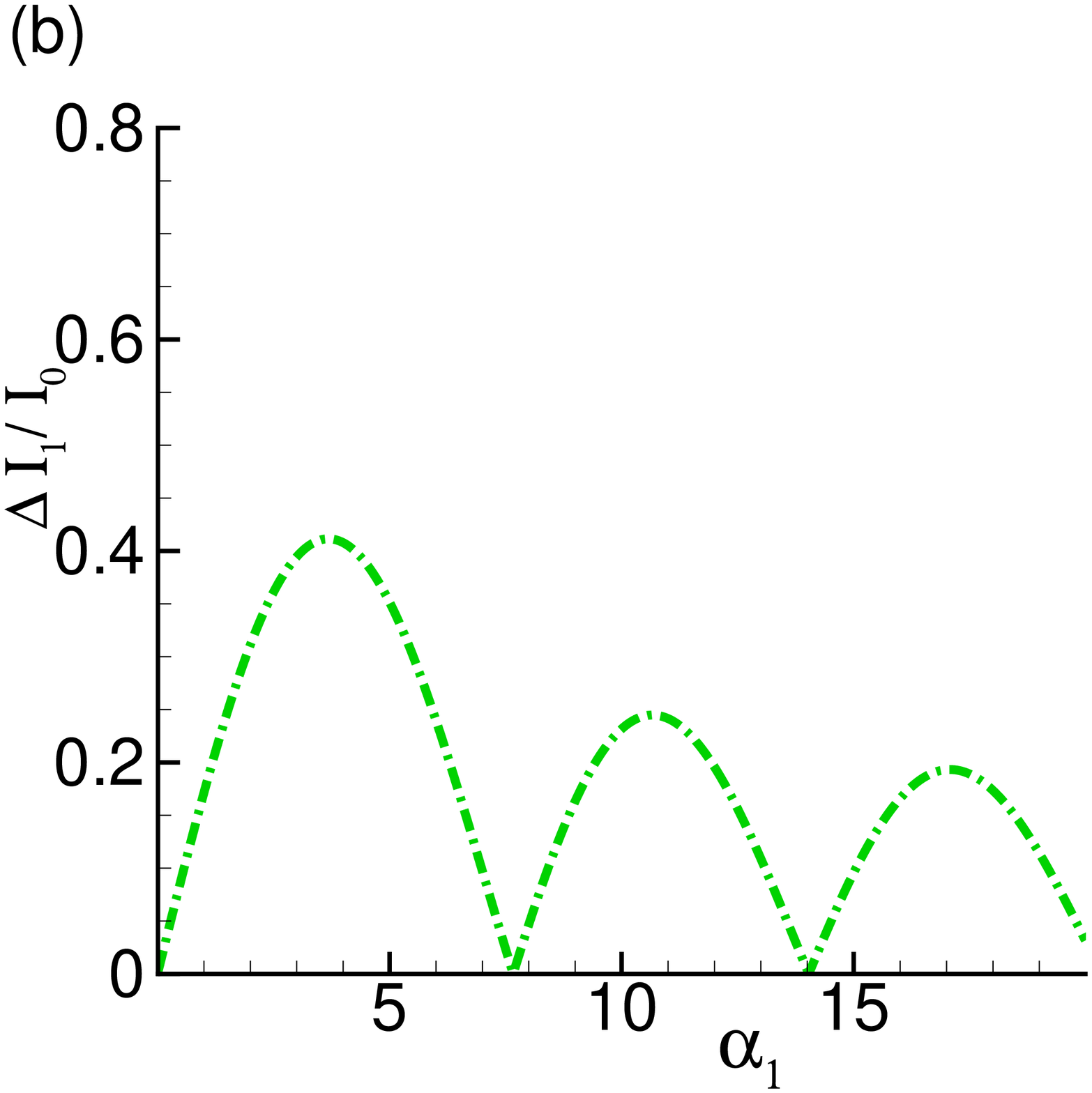}\\
\includegraphics[width=4.7cm]{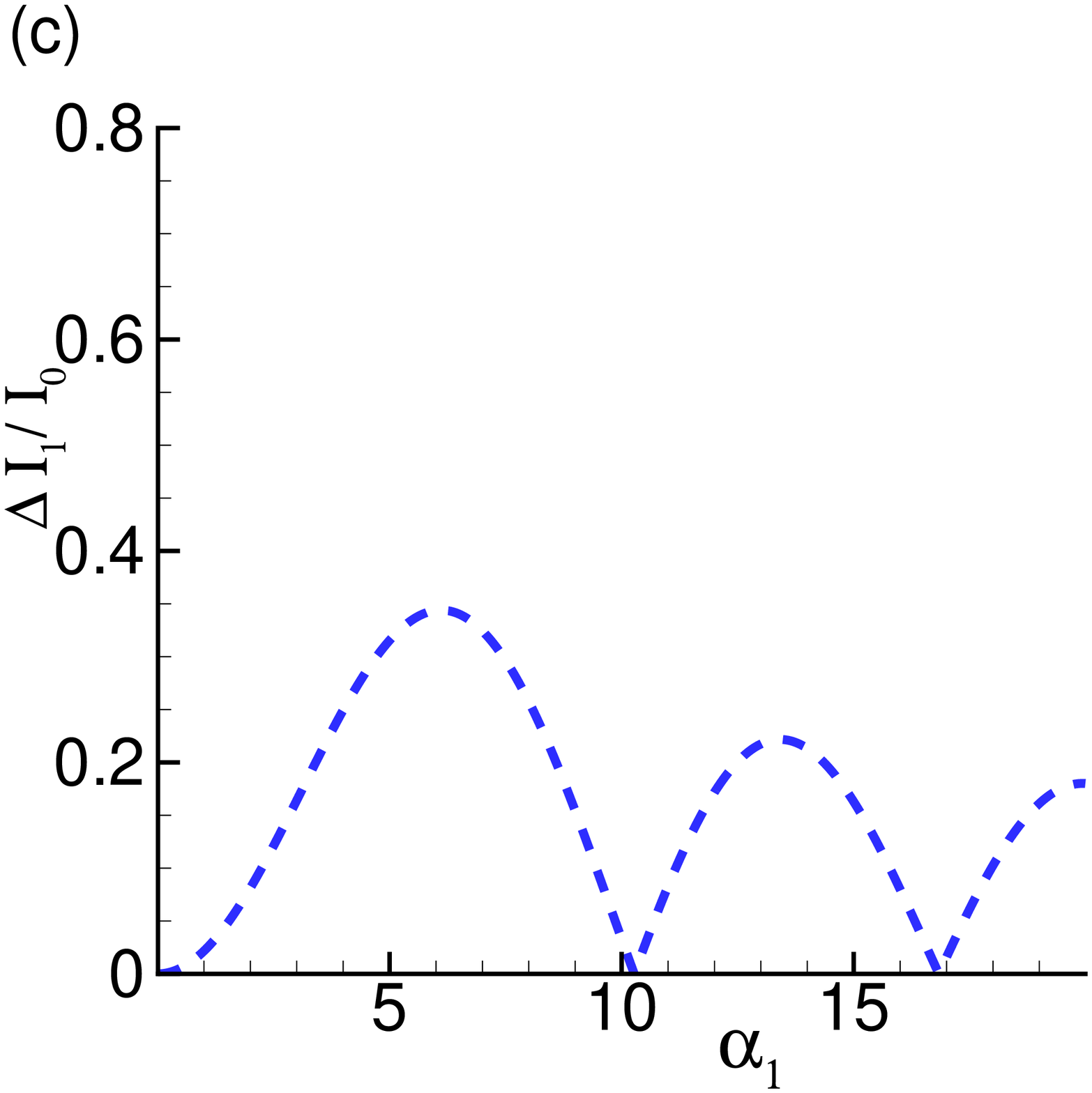}\includegraphics[width=4.7cm]{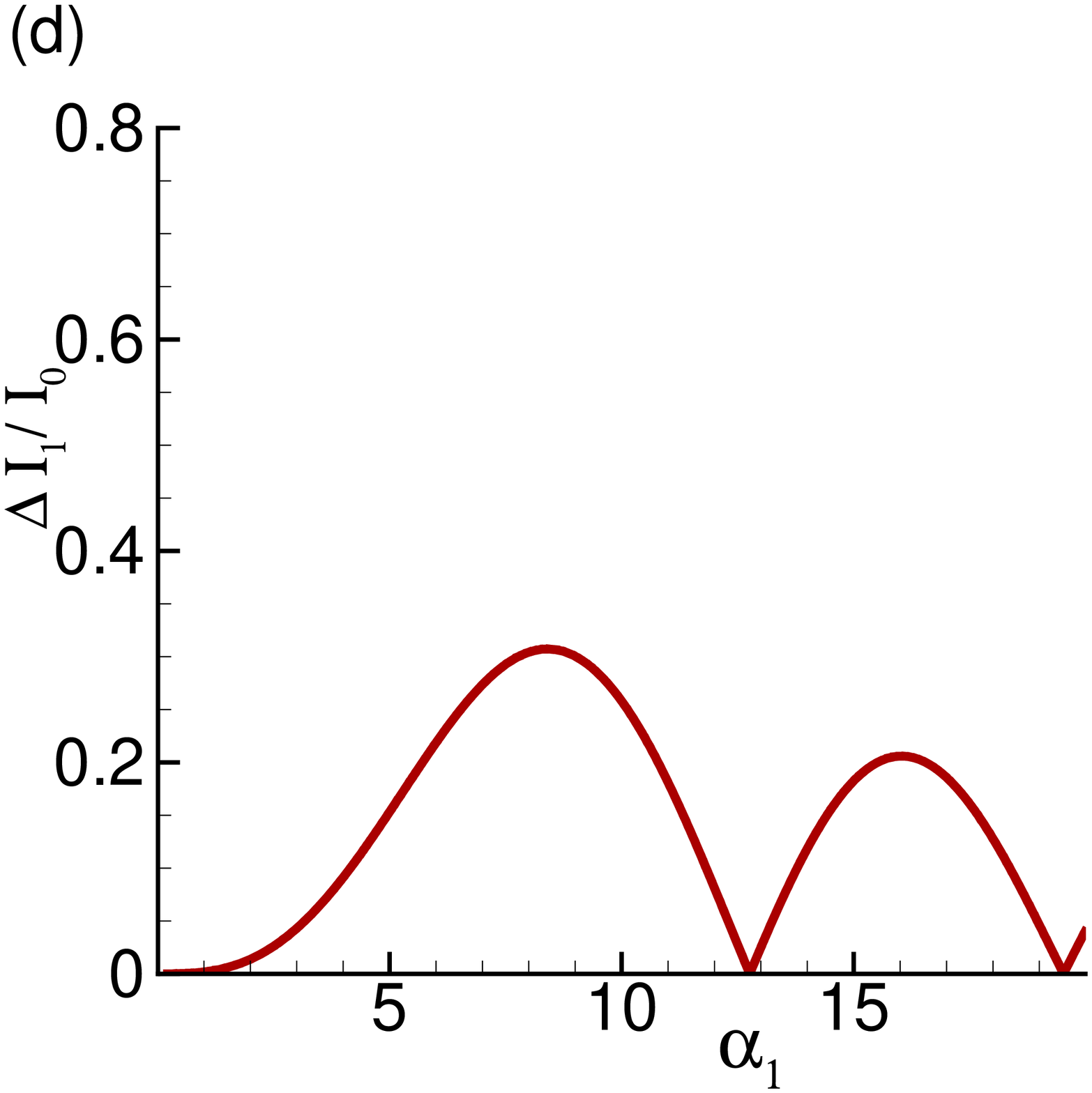}
\caption{The dependence of the Shapiro step width for $p$-wave JJs
at $\alpha = {\bf B}=h=0$ on the amplitude of the $ac$-voltage
$\alpha_1= 2eV_1/(\hbar \omega)$  at (a) $n_0=0$, (b) $n_0=1$, (c)
$n_0=2$, and (d) $n_0=3$. $D$ is chosen to be $D=0.5$.} \label{53}
\end{figure}

In this section, we compute the AC Josephson effect for the tunnel
junctions mentioned above. If there is the voltage in Josephson
junction $V(t)=V_0+ V_1 \cos{\omega t}$, then from Josephson
relation $\dot{\phi}=2eV/\hbar$ we get
\begin{eqnarray}
\phi(t) &=&  \varphi_0 + \omega_J t + \alpha_1 \sin{\omega t},
\label{phaseeq1}
\end{eqnarray}
where $\omega_J= 2eV_0/\hbar$ and $\alpha_1= 2eV_1/(\hbar \omega)$.
To obtain the AC Josephson current for such a voltage-biased JJ, we
use the following procedure. We consider a JJ with phase difference
$\phi$ and having Andreev bound state energies $E_n [\phi,\theta;
\alpha, h, B]$. The Josephson current at $T=0$ can be obtained from
these bound states by using $ I_J = (2e/\hbar) \sum_n \partial E_n
/\partial \phi \, \theta(-E_n)$. One can then obtain the AC
Josephson current by the replacement
\begin{eqnarray}
I_J^{\rm AC} = I_J[\phi \to \phi(t)] \label{acjos1}
\end{eqnarray}
For example, for pristine $p-$ wave JJs with $\alpha=h=B=0$, where
$E= \pm \Delta \sqrt{D} \cos(\phi/2)$ according to Eq.\ (\ref{ksy}),
this procedure leads to
\begin{eqnarray}
I_{J1}^{\rm AC} &=& \frac{I_0}{2} \sqrt{D} \sum_n J_n(\alpha_1/2)
\sin[(\varphi_0 +(\omega_J-2n \omega)t)/2] \nonumber\\ \label{ac1}
\end{eqnarray}
where $I_0= 2 e \Delta/\hbar$ and we have used the identity $\exp[i
a \sin \theta]= \sum_n J_n[a] \exp[i n \theta]$. Eq.\ (\ref{ac1})
reflects the fact that for $p$-wave junction one has Shapiro steps
at $\omega_J/\omega_n = 2n$ for integer $n$; the odd Shapiro steps
are absent. The width of the step corresponding to $n=n_0$ is given
by
\begin{eqnarray}
\Delta I_1 &=& \left|I_0 \sqrt{D} J_{n_0}(\alpha_1/2)
\right|\label{shap1}
\end{eqnarray}
where we have used the fact that the maximum width of the step
occurs at $\varphi_0 = \varphi_0^{m} = \pi$. The dependence of the
step width on $\alpha_1$ for different values of $n_0=0, 1, 2, 3$ is
presented in Fig. \ref{53}.

\begin{figure}[t]
\includegraphics[width=0.48 \linewidth]{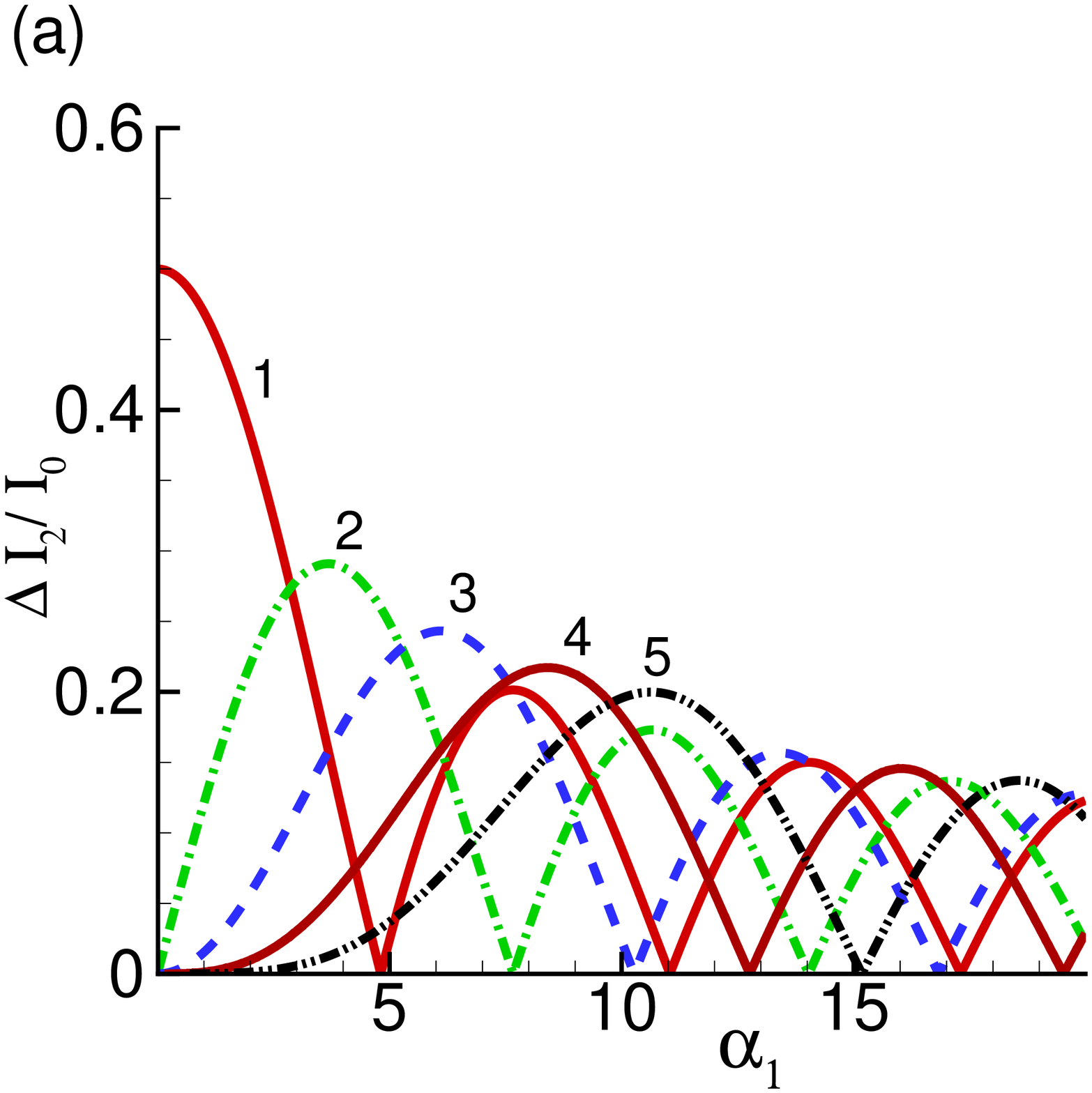}
\includegraphics[width=0.48 \linewidth]{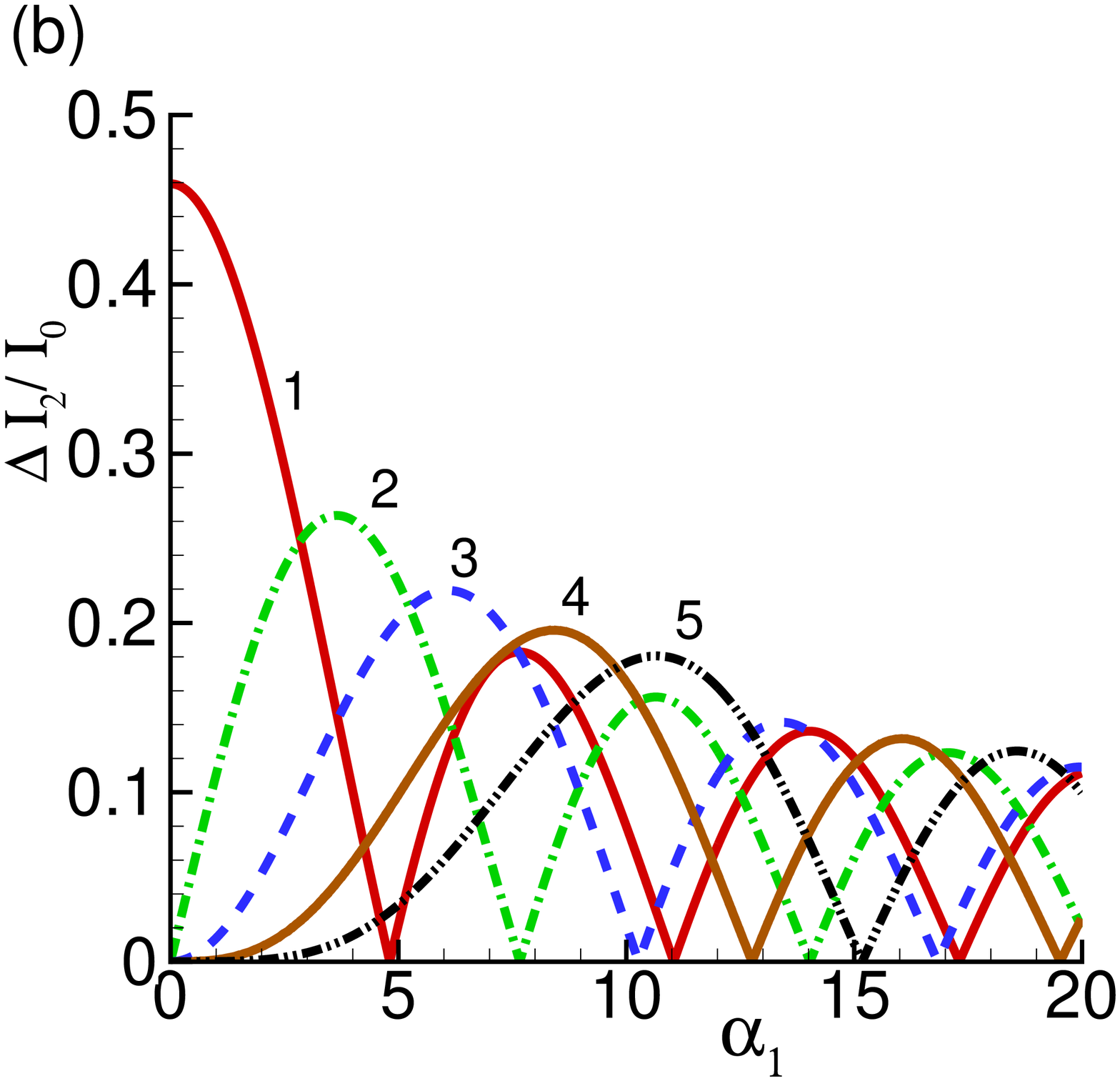}
\caption{The dependence of the Shapiro step width for $p$-wave JJs
at ${\bf B}=h=0$ and $\alpha \neq 0$ on the amplitude of the
$ac$-voltage $\alpha_1= 2eV_1/(\hbar \omega)$ according to
Eq.(\ref{ac2})  for two different values of the parameter $D_0$, (a)
$D_0=1/8<1/6$ at $D=0.5$ and $\beta =\sqrt{0.5}$, corresponding to
$\phi_0^m=\pi$  and (b) $D_0=7/40>1/6$ at $D=0.5$ and $\beta
=\sqrt{0.3}$. The curves in each figure are drawn by (1) solid red
curve at $n_0=0$, (2) green dot-dashed curve  at $n_0=1$, (3) blue
dashed curve  at $n_0=2$, (4) brown dotted curve at $n_0=3$, and (5)
black double dot-dashed curve at $n_0=4$.} \label{56}
\end{figure}

Next, we apply this procedure for the case where $B=h=0$ but $\alpha
\ne 0$. The Andreev bound state energy is given by Eq.\
(\ref{E-B=01}) and consists of four branches, i.e. each electron and
hole branch is split into two states. One of these split states,
corresponding to the $+$ sign (Eq. \ref{E-B=01}) are independent of
$\alpha$. For these two states, the AC Josephson current can be
easily shown to be given by Eq.\ (\ref{ac1}); the corresponding
Shapiro step width is given by Eq.\ (\ref{shap1}). In contrast, the
energy dispersion of the other two branches with designated by $-$
sign (Eq.\ \ref{E-B=01}) depend on the ratio $\alpha/v_F$ and can be
rewritten as
\begin{eqnarray}
E_-^s &=& \frac{s \Delta_0 \beta \sqrt{D}
\cos(\phi/2)}{\sqrt{1-D(1-\beta^2) \cos^2(\phi/2)}} \label{balpha1}
\end{eqnarray}
where $\beta= (1-\alpha/v_F)/(1+\alpha/v_F)$. We note that these
branches do not contribute to the Josephson current if $\alpha=v_F$.
For $\alpha < v_F$ the contribution from the $s=-$ branch to the
current is given by,
\begin{widetext}
\begin{eqnarray}
I_{J2}^{\rm AC} &=& \frac{I_0 \beta \sqrt{D}}{2} \frac{ \sum_n
J_n(\alpha_1/2) \sin[(\varphi_0 +(\omega_J-2n \omega)t)/2]}{\left
\{1 - D_0 - D_0 \sum_n J_n(\alpha_1) \cos[\varphi_0 +(\omega_J-n
\omega)t]\right\}^{3/2}} \label{ac2}
\end{eqnarray}
\end{widetext}
where $D_0= D(1-\beta^2)/2$. The Shapiro step width corresponding to
$n=n_0$ is given by
\begin{eqnarray}
\Delta I_2 &=& I_0 \sqrt{D} \beta \left|\frac{J_{n_0}[\alpha_1/2]
\sin(\phi_0^m/2)}{\left\{1-D_0 -D_0 J_{n_0}(\alpha_1) \cos(\phi_0^m)
\right\}^{3/2}} \right| \nonumber\\ \label{shap2}
\end{eqnarray}
In order to determine $\phi_0^m$, we need to find the value of
$\varphi_0$ which maximizes the Shapiro step width. This can be
computed easily from Eq.\ (\ref{ac2}) by maximizing the current
after setting $\omega_J=2 n_0 \omega$. This procedure yields
\begin{eqnarray}
\phi_0^m &=& 2 \arcsin \left[\sqrt{(1-2D_0)/(4 D_0)}\right] \quad
{\rm if} \,
\, D_0 > 1/6  \nonumber\\
&=& \pi  \quad {\rm otherwise}  \label{maxang1}
\end{eqnarray}
The dependence of the Shapiro step width on $\alpha_1$ (Eq.
(\ref{shap2})) is plotted, using Eq.\ (\ref{maxang1}) in Fig.
(\ref{56}) for $D_0<1/6$ (Fig.\ \ref{56}(a)) and $D_0>1/6$ (Fig.\
\ref{56}(b)). We note that Eqs.\ (\ref{shap2}) and (\ref{maxang1})
allow one to obtain the strength of the spin-orbit coupling in these
JJs from the Shapiro step width.

Next we consider the case where $\alpha=B=0$ and $h \ne 0$. Here the
energy dispersion corresponds to four branches as can be seen from
Eq.\ (\ref{E-B=02}). We first consider the branches corresponding to
$E_{\uparrow \downarrow}^-$. Here we note that when $E_{\uparrow
\downarrow}^-(s=\pm)<0$, the Josephson currents from these two
branches cancel each other. Similarly, if $E_{\uparrow
\downarrow}^-(s=\pm)> 0$, none of the branches contribute at $T=0$.
Thus these branches contribute to the Josephson current for
$\sqrt{D} \sqrt{\Delta^2-4 h^2} |\cos(\varphi/2)|> h$; the
presence/absence of Josephson currents from these branches can
therefore be used to estimate $D$ in these junctions, provided
$\Delta$ and $h$ are known. The Josephson current from these
branches and the corresponding Shapiro step widths (when the
above-mentioned condition is satisfied) are given by Eq.\
(\ref{ac1}) and (\ref{shap1}) respectively. In contrast, the
contribution from the branches corresponding to $E_{\uparrow
\downarrow}^+(s)$ are given by
\begin{widetext}
\begin{eqnarray}
I_{J3}^{\rm AC}/I_0 &=& -\frac{2e}{\hbar} \left[D h \sin \phi(t) +
\sum_{s=\pm} \frac{s \sqrt{D} (\Delta^2-4h^2 +4Dh(1+\cos\phi(t)))
\sin [\phi(t)/2]}{\left[ \Delta^2-4 h^2 + 2hD (1+
\cos(\phi(t)/2))\right]^{1/2}} \theta(-E_{\uparrow \downarrow}^-(s))
\right] \label{ac3}
\end{eqnarray}
\end{widetext}
We note that when both $E^-_{\uparrow \downarrow}(s=\pm)<0$, the
Josephson current is purely $2 \pi$ periodic and is given by
\begin{eqnarray}
I_{J4}^{\rm AC} &=&  -\frac{I_0 D h}{\Delta} \sum_n J_n(\alpha_1)
\sin[\varphi_0 + (\omega_J-n \omega)t] \label{ac4}
\end{eqnarray}
In this case one has both odd and even Shapiro steps with the step
width $\Delta I_4[n_0] =  |I_0 D h J_{n_0}(\alpha_1)/\Delta|$.
However, for $E_{\uparrow \downarrow}^-(s=+) >0$ and $E_{\uparrow
\downarrow}^-(s=-) < 0$, one has
\begin{widetext}
\begin{eqnarray}
I_{J5}^{\rm AC} &=& I_{J4}^{\rm AC} + \frac{\sqrt{D} [\Delta^2-4h^2
+4Dh +4Dh \sum_n J_n(\alpha_1)\cos(\phi_0+(\omega_J-n \omega)t)]
\sum_n J_n(\alpha_1/2) \sin[(\phi_0+(\omega_J-2n
\omega)t)/2]}{\left[ \Delta^2-4 h^2 + 2hD +2hD \sum_n J_n(\alpha_1)
\cos(\phi_0 +(\omega_J-n \omega)t)\right]^{1/2}} \nonumber\\
\label{ac5}
\end{eqnarray}
\end{widetext}
We note that in this case the JJ will have both $2 \pi$ periodic and
$4 \pi$ periodic components. The corresponding Shapiro step width is
given for $n=n_0$
\begin{widetext}
\begin{eqnarray}
\Delta I_5 &=& \Delta I_4[n_0]  \quad  n_0= 2m_0+1  \label{shap5} \\
&=& \Delta I_4[n_0] \left|\sin[\varphi_0^m]\right| +
\left|\frac{\sqrt{D} [\Delta^2-4h^2 +4Dh +4Dh J_{n_0}(\alpha_1)\cos
\phi_0 ] J_{n_0}(\alpha_1/2)\sin \phi_0/2}{\left[ \Delta^2-4 h^2 +
2hD +2hD J_{n_0}(\alpha_1) \cos\phi_0\right]^{1/2}} \right| \quad
n_0=2m_0 \nonumber
\end{eqnarray}
\end{widetext}
where $m_0$ is an integer and $\varphi_0^m$ denotes the value of
$\varphi_0$ for which the stepwidth is maximum. This value needs to
be numerically determined for the present case by minimizing Eq.\
(\ref{ac5}) at $\omega_J= 2 n_0 \omega$ with respect to $\varphi_0$.

Finally, we treat the case $B,h \ne 0$ and $\alpha=0$. Here the
Andreev bound states are given by Eq.\ (\ref{Eup-down1}) and
(\ref{Eup-up1}). For $E\equiv E_{\uparrow \downarrow}^{\pm}(s)$
(Eq.\ (\ref{Eup-down1})), there are four branches. For
$E=E^+_{\uparrow \downarrow}$ both the branches are above the Fermi
energy if $\sqrt{B^2 +h^2} \ge \Delta \sqrt{D}$. In this case, there
is no Josephson current contribution from these branches. Similarly,
for $E=E^-_{\uparrow \downarrow}$, the same condition results in
both the branches being below the Fermi level. In this case, their
contribution to the Josephson currents cancel each other. Thus the
contribution to the Josephson current from $E_{\uparrow \downarrow}$
occurs only when $\sqrt{B^2+h^2} < \Delta \sqrt{D}$. However, even
in this case, the contribution to the Josephson currents from the
positive ($E^+$) and negative ($E^-$) branches cancel each other and
the net Josephson current vanishes. A similar results can be easily
deduced for $E= E_{\uparrow \uparrow}^{\pm}$ (Eq.\ (\ref{Eup-up1}))
and $E_{\downarrow \uparrow}$ (Eq.\ (\ref{E1})) branches.

\section{Conclusion}
\label{sec6}

In this paper we study the Josephson current between 1D nanowires of
$p$-wave superconductors separated by an insulating barrier in the
presence of Rashba SOI and the magnetic fields ${\bf B}$ and $h$.
The presence of the SOI and Zeeman magnetic fields enlarges the
standard two-component $BdG$ system equations  to four-component
system equations (\ref{Sch1})- (\ref{Sch4}) for new BdG wave vector
$\eta_a(x)=\left(\eta_{a,\uparrow, +}^{\dag}(x), \eta_{a,
\downarrow,+}^{\dag}(x), \eta_{a, \downarrow, -} (x),
\eta_{a,\uparrow, -}(x)\right)$. The BdG equations
(\ref{Sch1})-(\ref{Sch4}) coincide with the standard BdG equations
in the absence of the in-plane magnetic field ${\bf B}$, which
provide only one relation  $\eta^{\ast}_{a, \sigma, b}/\eta_{a,
{\bar \sigma}, {\bar b}}$ between quasi-particle and quasi-hole
states, where $\sigma = \uparrow, \downarrow$ and ${\bar
\sigma}=\downarrow, \uparrow$. Instead, the BdG equations
(\ref{Sch1})-(\ref{Sch4}) in the presence of ${\it in-plane}$
magnetic field ${\bf B}$ and Rashba SOI provide three relations
$\eta^{\ast}_{a, \sigma, b}/\eta_{a,{\bar \sigma},{\bar b}}$,
$\eta^{\ast}_{a, \sigma, b}/\eta_{a,\sigma, {\bar b}}$, and
$\eta_{a, \sigma,b}/\eta_{a,{\bar \sigma}, b}$ between the
quasi-particle and quasi-hole states, corresponding to new Andreev
scattering channels. We studied in this paper all these scattering
channels in detail by generalizing the method of Ref.\
\onlinecite{ksy04} for study of Josephson junction with
$\delta$-function insulator between two $p$-wave superconductors to
systems with SOI and Zeeman fields. We have shown in this paper that
$\pi$-state is realized in Josephson junction with $p$-wave
superconductors. Moreover, we have demonstrated the existence of
magneto-Josephson effect in these systems. We note that although the
existence of the magneto-Josephson effect in a topological
superconductor has been predicted recently \cite{jpar13, kss12,
pjpa13}, the question of whether this effect is observable in
superconducting junctions with $p$-wave superconductors and the
presence of SOI was not addressed before. We have predicted in the
paper new Andreev-type tunneling channels for quasi-particles and
quasi-holes which are responsible to the magneto-Josephson effect.

In conclusion, we have studied Josephson effect in a junction between
two $p$-wave 1D nanowires in the presence of SOI and Zeeman fields. We have
analyzed the Josephson current in these junctions and provided
analytical expressions of the Andreev bound states in several
limiting cases. We have also demonstrated the presence of
magneto-Josephson effect in these junctions.
Our heoretical predictions are shown to be verifiable by
straightforward experiments on these systems.

\section*{Acknowledgments}
The authors kindly acknowledge the Scientific Fund of State Oil
Company of Azerbaijan Republic (SOCAR) for financial support of
2019-2020 grant entitled 'Study of the impurity and correlation
effects in graphene, fullerene and other topological
nanostructures'. The reported study was partially funded by the RFBR
research projects 18-02-00318 and 18-52-45011-IND. Part of the
numerical calculations were made in the framework of the RSF project
18-71-10095. KS thanks DST, India for support through project
INT/RUS/RFBR/P-314.

\appendix
\section{Andreev bound state energies at ${\bf B} \neq 0$}

Andreev bound state energies are obtained from the conditions $F_{\uparrow \downarrow}^{\ast}(E_{\uparrow \downarrow})=0$,
$F_{\uparrow \uparrow}^{\ast}(E_{\uparrow \uparrow})=0$, $F_{\uparrow \downarrow}({\tilde E}_{\uparrow \downarrow})=0$ corresponding to three
channels, and also from the conditions, obtained by interchanging the spin orientations as
 $F_{\downarrow \uparrow}^{\ast}(E_{\downarrow \uparrow})=0$, $F_{\downarrow \downarrow}^{\ast}(E_{\downarrow \downarrow})=0$,
$F_{\downarrow \uparrow}({\tilde E}_{\downarrow \uparrow})=0$.

Andreev bound state energy $E_{\uparrow \downarrow}$ is obtained from the condition $F_{\uparrow \downarrow}^{\ast}(E_{\uparrow \downarrow})=0$
with the expression (\ref{Fup-down}) yielding
\begin{widetext}
\begin{eqnarray}
&&\left\{(E_{\uparrow \downarrow}+h)\left[(E_{\uparrow \downarrow}-h)^2 +k^2(v_F+\alpha)^2+B^2-\Delta^2\right]^2-
2B^2E_{\uparrow \downarrow} \right\}^2\left(1-D\cos^2\frac{\varphi}{2}\right)-\nonumber\\
&&\left\{k(v_f-\alpha)\left[(E_{\uparrow \downarrow}-h)^2+k^2(v_F+\alpha)^2+B^2-\Delta^2\right]+2kB^2\alpha\right\}^2
D\cos^2\frac{\varphi}{2}=0.
\label{AEup-down}
\end{eqnarray}
This expressin depends apart from the parameters $\alpha$, ${\bf B}$, and $h$ also on the momentum $k^2$.
Expression for $k^2_{\pm}$, obtained from the energy spectrum (\ref{Eo}), reads
\begin{eqnarray}
&&k^2_{\pm}(v_F^2-\alpha^2)^2=(E^2-h^2-B^2-\Delta^2)(v_F^2-\alpha^2)-2(Ev_F+\alpha h)^2+2\Delta^2v_F^2 \pm
\Big\{\big[(E^2-h^2-B^2-\Delta^2)(v_F^2-\alpha^2)-\nonumber\\
&& 2(Ev_F+\alpha h)^2+2\Delta^2v_F^2\big]^2-
\left[(E^2-h^2-B^2-\Delta^2)^2-4\Delta^2(h^2+B^2)\right](v_F^2-\alpha^2)^2 \Big\}^{1/2}.
\label{k2}
\end{eqnarray}
Elimination of $k^2$,  by substituting it from (\ref{k2}) into  Eq. (\ref{AEup-down}), yields a general
expression to find $E_{\uparrow \downarrow}$, which is not easy to solve exactly.

The condition $F_{\uparrow \uparrow}^{\ast}(E_{\uparrow \uparrow})=0$ with the expression (\ref{Fup-up}) yields
\begin{equation}
k^2(Ev_F+\alpha h)^2-\Delta^2(E^2_{\uparrow \uparrow}+\alpha^2 k^2)+\Delta^2 D(E^2_{\uparrow \uparrow}+\alpha^2 k^2)
\cos^2 \frac{\varphi -\theta}{2}=0.
\label{AEup-up2}
\end{equation}
Routine calculations, after substitution of $k^2$ from Eq. (\ref{k2}) to this equation, result in
\begin{eqnarray}
&&\left\{(v_F^2- \alpha^2)E^2_{\uparrow \uparrow}\Delta^2\left(1-D\cos^2\frac{\varphi - \theta}{2}\right)-(E^2_{\uparrow \uparrow}
+\Delta^2-B^2-h^2) \left[(E_{\uparrow \uparrow}v_F+\alpha h)^2- \alpha^2 \Delta^2\left(1-D\cos^2 \frac{\varphi -
\theta}{2}\right)\right]\right\}^2-\nonumber\\
&&4E^2_{\uparrow \uparrow} \Delta^2 D\cos^2\frac{\varphi - \theta}{2} (E_{\uparrow \uparrow}v_F +\alpha h)^2
\left[(E_{\uparrow \uparrow} v_F +\alpha h)^2-\alpha^2 \Delta^2 \left(1-D \cos^2 \frac{\varphi - \theta}{2}\right)\right]=0
\label{AEup-up3}
\end{eqnarray}
This equation can be solved numerically for a general case when $\alpha,~ {\bf B},~ h \neq 0$.
The expression for $F_{\downarrow \downarrow}^{\ast}(E_{\downarrow \downarrow})=0$ is obtained from Eq. (\ref{AEup-up3})
by interchanging $\alpha \to -\alpha$, $h \to - h$, and $\theta \to - \theta$.

The bound state energy in the particle-particle channel with opposite spin orientations is determined from the
condition $F_{\uparrow \downarrow}(E'_{\uparrow \downarrow})=0$, which can be written by using the expression (\ref{FAup-down})
for $F_{\downarrow \uparrow}$ as,
\begin{equation}
\alpha^2 k^2- \left(E^{' 2}_{\uparrow \downarrow}+\alpha^2 k^2\right)D\sin^2\frac{\theta}{2}=0.
\label{FAup-down1}
\end{equation}
Substituting $k^2$ from Eq. (\ref{k2}) to this equation yields the equation to determine $E'_{\uparrow \downarrow}$,
\begin{eqnarray}
&&\alpha^4 \left[ (E_{\uparrow \downarrow}^{' 2}+ \Delta^2 -B^2-h^2)^2-4 \Delta^2 E^{' 2}_{\uparrow \downarrow} \right]+2\alpha^2 D \sin^2
\frac{\theta}{2} \Big\{v_F^2 E^{' 4}_{\uparrow \downarrow} + (v_F^2-\alpha^2)\Delta^2 E^{' 2}_{\uparrow \downarrow}+
(v_F^2+\alpha^2)B^2 E^{' 2}_{\uparrow \downarrow}+\nonumber\\
&& (v_F^2+3\alpha^2) h^2 E^{' 2}_{\uparrow \downarrow}-4v_F\alpha h E^{' 3}+\alpha^2(\Delta^2-B^2-h^2)^2 \Big\}+D^2\sin^4\frac{\theta}{2}
\Big\{ v_F^2(v_F^2-4\alpha^2)E^{' 4}_{\uparrow \downarrow}+\nonumber\\
&& 2\alpha^2 E^{' 2}_{\uparrow \downarrow}\left[v_F^2\Delta^2-v_F^2B^2-(v_F^2+2\alpha^2)h^2\right]-8\alpha^3 v_FhE^{' 3}_{\uparrow \downarrow}+
\alpha^4 (\Delta^2-B^2-h^2)^2\Big\}=0.
\label{FAup-down2}
\end{eqnarray}
Note that the condition $F_{\downarrow \uparrow}(E'_{\downarrow \uparrow})=0$ provides for $E'_{\downarrow \uparrow}$ exactly the same
expression as (\ref{FAup-down2}), i. e. $E'_{\downarrow \uparrow}=E'_{\uparrow \downarrow}$. Indeed, a reality of this fact can
be tested according to the rule that interchanging the spin directions is equivalent to the replacements of
$\alpha \to - \alpha$, $h \to - h$, and $\theta \to - \theta$.
\end{widetext}

%

\end{document}